
\input harvmac
\input epsf
%
%
%
%
%

\parindent=0pt
\Title{SHEP 95-21}{\vbox{\centerline{Momentum Scale Expansion}\vskip2pt
\centerline{of}\vskip2pt\centerline{Sharp Cutoff Flow Equations}}}

\centerline{\bf Tim R. Morris}
\vskip .12in plus .02in
\centerline{\it Physics Department}
\centerline{\it University of Southampton}
\centerline{\it Southampton, SO17 1BJ, UK}
\vskip .7in plus .35in

\centerline{\bf Abstract}
\smallskip 
We show how the exact renormalization group for the  effective action
with 
a sharp momentum cutoff, may be organised by expanding one-particle
irreducible parts in terms of homogeneous
functions of momenta of integer degree (Taylor expansions not being possible).
A systematic series of approximations -- the $O(p^M)$ approximations --
result from discarding from these parts,
all terms of higher than the $M^{\rm th}$ degree.
These approximations preserve a field reparametrization invariance, ensuring
that the field's anomalous dimension is unambiguously determined.
The lowest order approximation coincides with the
local potential approximation to the Wegner-Houghton equations.
We discuss the practical difficulties with extending the approximation
beyond $O(p^0)$.


\vskip -1.5cm
\Date{\vbox{
{hep-th/9508017}
\vskip2pt{August, 1995.}
}
}
\catcode`@=11 
\def\slash#1{\mathord{\mathpalette\c@ncel#1}}
 \def\c@ncel#1#2{\ooalign{$\hfil#1\mkern1mu/\hfil$\crcr$#1#2$}}
\def\lsim{\mathrel{\mathpalette\@versim<}}
\def\gsim{\mathrel{\mathpalette\@versim>}}
 \def\@versim#1#2{\lower0.2ex\vbox{\baselineskip\z@skip\lineskip\z@skip
       \lineskiplimit\z@\ialign{$\m@th#1\hfil##$\crcr#2\crcr\sim\crcr}}}
\catcode`@=12 

\def\phi{\varphi}

\def\te#1{\theta_\epsilon( #1,\Lambda)}
\def\epsilon{\varepsilon}
\def\p{{\bf p}}
\def\P{{\bf P}}
\def\Q{{\bf Q}}
\def\q{{\bf q}}
\def\r{{\bf r}}
\def\x{{\bf x}}
\def\y{{\bf y}}
\def\tr{{\rm tr}}
\def\D{{\cal D}}
\def\E{{\cal E}}

\def\ins#1#2#3{\hskip #1cm \hbox{#3}\hskip #2cm}
\def\frac#1#2{{#1\over#2}}

\lref\lparef{J.F. Nicoll et al, Phys. Rev. A13 (1976)
1251, {ibid} A17 (1978) 2083;\quad
K. Kawasaki et al, in
    ``Perspectives in Statistical Physics'', ed. H. Ravech\'e  (1981)
          North-Holland;\quad
A. Parola et al, Phys. Rev. A31 (1985) 3309,  Phys. Rev. E48 (1993) 3321;\quad
A. Meroni et al, Phys. Rev. A42 (1990) 6104;\quad
R. Lipowsky and M.E. Fisher, Phys. Rev. B36 (1987) 2126;\quad
P. Hasenfratz and J. Nager, Z. Phys. C37 (1988) 477;\quad
A. Margaritis et al, Z. Phys. C39 (1988) 109;\quad
M. Maggiore, Z. Phys. C41 (1989) 687;\quad
C. Bagnuls and C. Bervillier, Phys. Rev. B41 (1990) 402;\quad
See the review and refs therein:\quad
         T.S. Chang 
         et al, Phys. Rep. 217 (1992) 279;\quad
T.E. Clark et al, Nucl. Phys. B402 (1993)
   628, Phys. Rev. D50 (1994) 606, Phys. Lett. B344 (1995) 266;\quad
M. Alford and J. March-Russell, Nucl. Phys. B417 (1994) 527;\quad
K. Kimura et al,  Mod. Phys. Lett. A9 (1994) 2587;\quad
P.E. Haagensen  et al, Phys. Lett. B323 (1994) 330;\quad
S-B Liao et al, Phys. Rev. D51 (1995) 748, 4474, DUKE-TH-94-63,
hep-th/9404086;\quad
K. Halpern and K. Huang,  Phys. Rev. Lett. 74 (1995) 3526;\quad
J.R. Shepard et al, UC/NPL-1112, hep-lat/9412111.}
\lref\wegho{F.J. Wegner and A. Houghton, Phys. Rev. A8 (1973) 401.}
\lref\nico{J.F. Nicoll, T.S. Chang and H.E. Stanley,
           Phys. Rev. Lett. 33 (1974) 540.}
\lref\nicii{J.F. Nicoll, T.S. Chang and H.E. Stanley,
            Phys. Lett. 57A (1976) 7.}
\lref\pare{A. Parola and L. Reatto, Phys. Rev. Lett. 53 (1984) 2417.}
\lref\hashas{A. Hasenfratz and  P. Hasenfratz, Nucl. Phys. B270 (1986) 687.}
\lref\erg{T.R. Morris, Int. J. Mod. Phys. A9 (1994) 2411.}
\lref\deriv{T.R. Morris, Phys. Lett. B329 (1994) 241.}
\lref\trunc{T.R. Morris, Phys. Lett. B334 (1994) 355.}
\lref\twod{T.R. Morris, Phys. Lett. B345 (1995) 139.}
\lref\revi{T.R. Morris, in {\it Lattice '94}, Nucl. Phys. B(Proc. Suppl.)42
          (1995) 811.}
\lref\ui{T.R. Morris, SHEP 95-07, hep-th/9503225, to be published in Phys.
         Lett. B.}
\lref\alford{M. Alford, Phys. Lett. B336 (1994) 237.}
\lref\oldrepar{T.L. Bell and K.G. Wilson, Phys. Rev. B11 (1975) 3431\semi
E.K. Riedel, G.R. Golner and K.E. Newman, Ann. Phys. 161 (1985) 178.}
\lref\gol{G.R. Golner, Phys. Rev. B33 (1986) 7863.}
\lref\pol{J. Polchinski, Nucl. Phys. B231 (1984) 269.}
\lref\tetwet{N. Tetradis and C. Wetterich, Nucl. Phys. B422 (1994) 541.}
\lref\kogwil{K. Wilson and J. Kogut, Phys. Rep. 12C (1974) 75.}
\lref\red{F.J. Wegner, J. Phys. C7 (1974) 2098.} 
\lref\wein{Equivalent arguments, on a purely diagrammatic level, were given
  in: ``Critical Phenomena for Field Theorists'', S. Weinberg, lectures, Erice
  Subnucl. Phys. (1976) 1.}
\lref\zinn{See e.g.  J. Zinn-Justin,
            ``Quantum Field Theory and Critical Phenomena'' (1993)
             Clarendon Press, Oxford.}

\parindent=15pt

\newsec{Introduction.}
In ref.\refs{\wegho}, Wegner and Houghton formulated an exact
renormalization group equation for the Wilsonian effective
action\foot{a.k.a. Hamiltonian in statistical mechanics language}\
$S_\Lambda[\phi]$ where the associated momentum cutoff
$\Lambda$
is taken to be sharp: only momentum modes $p$ satisfying $p<\Lambda$ are
kept in the functional integral. The flow equation corresponds to computing
the induced change in $S_\Lambda$ as $\Lambda$ is lowered to
$\Lambda-\delta\Lambda$, by
integrating out  the modes with  $\Lambda-\delta\Lambda<p<\Lambda$.
 In the same
paper, the authors solved for the Wilson fixed point (i.e. massless case) of
the large $N$ limit of three dimensional $O(N)$
scalar field theory\foot{i.e. the large $N$ limit $N$-vector model,
equivalent\ref\stan{H.E. Stanley, Phys. Rev. 176 (1968) 718.}\
to the exactly solved spherical model\ref\james{G.S. Joyce, Phys. Rev. 146
(1966) 349.}.}, using the exact renormalization group language. This solution
involved approximating the effective action by just a
general effective potential $V_\Lambda(\phi)$:
\eqn\LPA{S_\Lambda[\phi]=\int\!\! d^D\!x\ \ \half(\partial_\mu\phi)^2+
V_\Lambda(\phi)\quad.}
In the large $N$ limit this approximation
effectively becomes exact. In ref.\refs{\nico},
Nicoll, Chang and Stanley
proposed to use such a {\sl local potential approximation} 
even when there is no justification in terms of the smallness of 
$1/N$.
\vphantom{\refs{\nico\nicii\pare\hashas\trunc\alford\lparef
\revi{--}\ui}}\def\tons{\refs{\nico{--}\ui}}$\!\!$
Such a local potential turns out to be
one-particle irreducible and may equivalently be regarded as
an approximation to the sharp cutoff flow equations for the
Legendre effective action\refs{\nicii,\erg}.\foot{a.k.a.
course-grained Helmholtz free energy. This argument is reviewed
later.}\
This approximation has proved to be very robust and has been
employed\foot{and sometimes rediscovered. For derivations see e.g.
refs.\refs{\wegho,\nico,\hashas,\trunc}.}\ many times since\tons.
\vphantom{\refs{\deriv,\twod,\gol,\tetwet}
\nref\genbyo{See for example:\quad
G.R. Golner, Phys. Rev. B8 (1973) 339;\quad
C. Wetterich, Phys. Lett. B301 (1993) 90 and refs therein;\quad
A.E. Filippov and A.V. Radievsky, Phys. Lett. A169 (1992) 195;\quad
R.D. Ball et al, Phys. Lett. B347 (1995) 80.}}$\!\!$
Obviously, it is very important to be able to establish such approximations,
because there are many situations in quantum field theory where no
genuinely small parameter 
 exists to control the approximation\ref\alll{See for example
refs.\refs{\nico{--}\genbyo}}.

One of the main reasons for the present paper is to provide further
theoretical justification for the local potential approximation by
demonstrating
that it may effectively be regarded simply as the start of a
certain expansion of $S_\Lambda$
in powers of momenta. This is by no means as trivial as it
might appear at first
sight. Quite apart from the fact that
a direct expansion in small momenta of the one-particle reducible parts of
$S_\Lambda$ vanishes to all orders, a Taylor expansion in momentum
components $p^\mu$ cannot any way be
implemented: the sharp cutoff in the flow equations
induces non-analyticity at the origin of
 momentum space,
reflecting the non-local behaviour 
in position space.  Instead, the expansion is performed via
the one-particle irreducible parts, and the best that
one can achieve is an expansion in {\sl momentum scale}\refs{\erg}
 $p\sim\sqrt{p^\mu p_\mu}$,
the $M^{\rm th}$
approximation resulting from dropping all terms beyond $O(p^M)$.
This raises the spectre
that the lowest order approximation -- $O(p^0)$, in which the momentum
scale dependence is completely discarded, nevertheless still
involves Green functions with non-trivial
dependence on the {\sl angles} between their momenta. It turns out
that this dependence
cancels however, and it is in this way that we recover the local potential
approximation\tons.

Our second main motivation is unfortunately
more negative. We wish to provide the evidence for the claim
in ref.\refs{\deriv}:
that the momentum scale expansion of the sharp cutoff flow equations
is not a practical technique, i.e. beyond $O(p^0)$,
because only certain truncations (of the field dependence)
are calculable in
closed form, and this is not good enough\refs{\trunc}.
It is not good enough because, in contrast to a momentum scale expansion
-- which might be expected to
work\foot{and certainly seems to so far\refs{\revi}}\
since it corresponds to an expansion
in (an appropriately defined) `localness' of the effective
Lagrangian, truncations of the field
dependence have limited accuracy and reliability\refs{\trunc}.
We want to emphasise here that our claim that there are
practical difficulties in extending the momentum scale expansion beyond
$O(p^0)$, merely amounts to an admission of our own lack of ability
in extending it in a substantive way. It would certainly be worthwhile
to  
improve on these attempts (presented in sect.6), because
a simple two-loop example -- and general arguments
(c.f. ref.\refs{\erg} and appendix B), suggest that the sharp
cutoff momentum scale expansion should yield the fastest
convergence (i.e. faster than
the smooth cutoff momentum expansions\refs{\deriv,\twod}).

Despite the increasing interest in  utilising the
exact renormalization group to develop non-perturbative
approximations\alll, there have so far
been relatively few works concerned with
going beyond the local potential
approximation\refs{\revi{--}\genbyo}. 
Part of the
point of sect.6 is also pedagogical: we provide  simple concrete
illustrations of the discussions in earlier sections, but also
reemphasise\refs{\revi,\deriv{--}\gol,\oldrepar}
in a simple context the important
issue of field reparametrization invariance. It has been
 understood in the condensed matter literature
for quite some time\refs{\oldrepar,\gol}
that it is important\foot{for linear renormalization group
transformations such as are exclusively discussed in the present
context\refs{\alll}}\
to preserve a field reparametrization invariance, e.g. simply
\eqn\resca{\phi\mapsto \phi/\lambda\quad,}
in the flow equations. For example, if  field reparametrization invariance
is broken by the approximation scheme,
the field's anomalous dimension can no longer be determined uniquely --
the value depends on at least one unphysical parameter.
Approximations by expansion in momenta (equivalently derivative expansion
in the smooth cutoff case) generally break reparametrization
invariance\refs{\gol}.
To date, only two forms of cutoff function are known that allow
a field reparametrization invariance after  approximation by momentum
expansion: the sharp cutoff studied
here, and power-law smooth cutoffs\refs{\deriv}.

Our formulation at first closely follows that of ref.\refs{\erg}, but we
will organise things differently. Therefore we feel it is preferable to give as
much as possible a self-contained account, indicating
the similarities or differences to ref.\refs{\erg} as they arise.
In sect.2, we rederive the sharp cutoff flow equations, taking due care of
the limit implied, but also incorporate explicitly the possibility of
an anomalous dimension, and display the field reparametrization invariance.
In sect.3, the momentum scale expansion, and the approximations which it
suggests, are defined -- and we stress again the fact that Taylor expansions
in momenta cannot be implemented. This means that the important property
of locality needs to be recovered, and we do this in sect.4; we also explain
why, and how, the uniqueness of the expansion then follows. In sect.5 we
work out the approximation to $O(p^0)$ and demonstrate that this coincides
with the local potential approximation\refs{\nico}. Sect.6 addresses by
example the new issues that arise in going beyond $O(p^0)$, computing
several examples, in particular
we indicate why we seem from the practical point of view
 to be limited to truncations of the field
dependence, and discuss -- with reference to a simple truncation --
the issue of field reparametrization invariance. Finally, in sect. 7 we
present our summary, making some general observations on the momentum
expansion presented here.

Before embarking on the paper proper, we review below
 the arguments that lead to expressing the flow
in terms of a Legendre effective action $\Gamma_\Lambda[\phi]$
with I.R. (infra-red) cutoff $\Lambda$.
The reason that this, or an equivalent formulation, must be used is that
 the Polchinski (or equivalently\refs{\deriv} the
 Wilson) effective action $S_\Lambda[\phi]$, for the effective
theory with U.V. (ultraviolet) cutoff $\Lambda$,
has a tree structure, composed of full propagators with I.R.
cutoff $\Lambda$, and one-particle irreducible parts which are those
generated by $\Gamma_\Lambda[\phi]$. This structure is already manifest in
the graphical form of the Polchinski\refs{\pol} (equivalently\refs{\deriv}
Wilson\refs{\kogwil}, or in the limit of sharp cutoff, Wegner's\refs{\wegho})
equation
for the vertices of $S_\Lambda$, which is displayed in \fig\pole{The
Polchinski
equation for the vertices of $S_\Lambda$.
The vertices are drawn as open circles.
In the limit of sharp cutoff, the
black dot represents a delta-function
restriction of the momentum $q$, in the propagator, to $q=
\Lambda$.}. (These trees take account of the purely
classical fluctuations that are integrated out as $\Lambda$ is reduced, and
would be there even in an effective action
for the classical field theory, while the one-particle irreducible parts
summarise the integrated out quantum field fluctuations).

\bigskip\medskip
\centerline{
\epsfxsize=\hsize\epsfbox{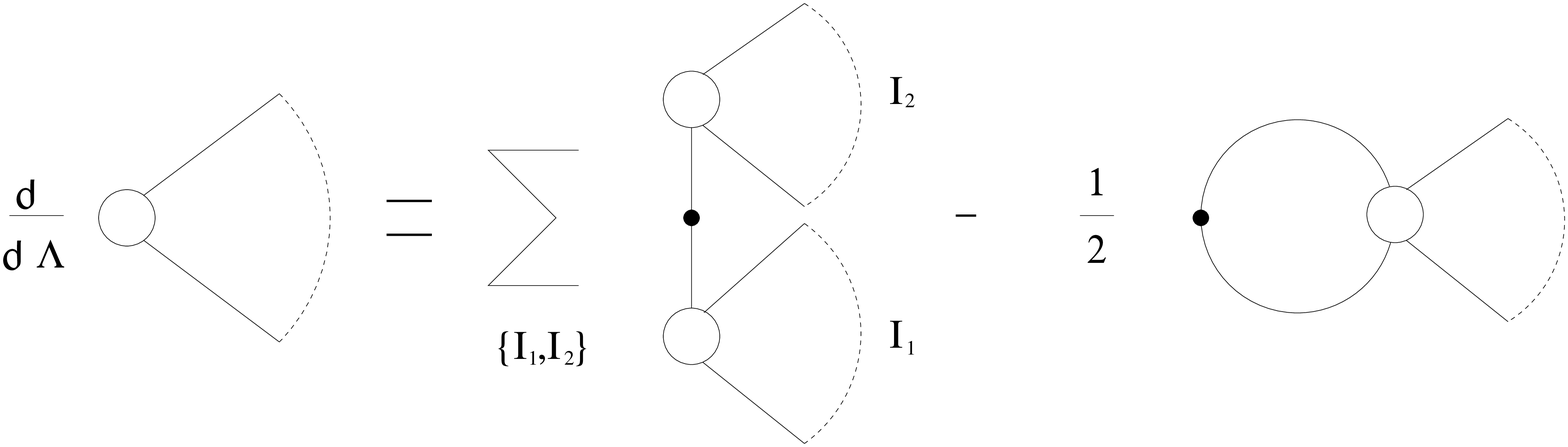}}  
\bigskip
\centerline{\vbox{\noindent {\bf Fig.1.} The Polchinski
equation for the vertices of $S_\Lambda$.
The vertices are drawn as open circles.
In the limit of sharp cutoff, the
black dot represents a delta-function
restriction of the momentum $q$, in the propagator, to $q=
\Lambda$.
}}
\bigskip\medskip

To resolve the limiting procedure
implicit in taking a sharp cutoff, it is necessary to take this structure
into account, which has the effect of reducing the equations to those
(or closely equivalent to those) for
$\Gamma_\Lambda[\phi]$ (c.f. \refs{\erg,\wein}).
Yet again, we must preserve the tree structure
on taking a momentum expansion. The
momentum expansion na\"\i vely
corresponds to Taylor expanding $S_\Lambda$ in the scale of
the external momenta (to the vertices of $S_\Lambda$), regarding this as small
compared to the cutoff $\Lambda$. In the sharp cutoff limit, this would cause
 all tree terms with internal propagators\foot{such as that of the first
term in \pole}\
to vanish however, since the internal propagator is furnished with a sharp
I.R. cutoff
$\Lambda$ and the momentum flowing through this propagator is of the same
scale as the external momenta by momentum conservation. Clearly
this is too great a mutilation of the theory, in particular
not only do all tree level corrections get discarded in the process,
but all loop diagrams
with more than one vertex get discarded also
(since these arise from substituting the tree parts
of $S_\Lambda$ into the second term in \pole). Instead we apply a
 momentum
expansion {\sl only to the one-particle irreducible vertices}, equivalently
to $\Gamma_\Lambda$.
In these cases all internal
propagators are integrated over momenta greater than $\Lambda$ (as follows
from integrating \pole\ between the overall U.V. cutoff $\Lambda^f_0$
and $\Lambda$), and
the momentum expansion corresponds to expanding the external
momenta, regarded as small compared to these internal momenta. (Actually
this oversimplifies a little: the
structure is hierarchical, with `inner' loops containing momenta larger than
those of the `outer' loops and inner propagators expanded in the outer loops
momenta\refs{\erg}). In this way no propagators are lost,
the diagrammatic structure is respected, and we might reasonably
hope that the momentum expansion leads to a convergent numerical\foot{It is not
in fact an expansion in a small parameter because the external
momenta $p$ are integrated out over the range $p<\Lambda$.}\ series.
The model two-loop calculation presented in ref.\refs{\erg} (and
reviewed in appendix B) is encouraging in
this respect since it results in a rapidly
convergent numerical series.

Finally we briefly mention an immediate consequence
of the tree structure: Setting the external momenta for the vertices of
$S_\Lambda$ to zero, kills all the tree terms. Therefore the Wilson
effective potential (supplied with U.V. cutoff $\Lambda$) coincides with
the effective potential in the Legendre effective action, if the latter
is computed with I.R. cutoff $\Lambda$ \refs{\erg}. This explains our statement
in the first paragraph.

\newsec{The flow equations.}
Thus we take as our starting point the partition function defined as
\eqn\zorig{\exp W_\Lambda[J]=\int\!\D\phi\
\exp\{-\half\phi.C^{-1}.\phi-S_{\Lambda^f_0}[\phi]+J.\phi\}\ \ ,}
regularised by an overall U.V. momentum cutoff $\Lambda_0^f$.
The notation is essentially the same as previously\refs{\erg,\deriv},
so two-point functions are often regarded as matrices in position or
momentum ($\q$)
space, one-point functions as vectors, and contractions indicated by a
dot. Momentum conserving $\delta$-functions are factored out when appropriate.
We work in $D$ Euclidean dimensions with a single real scalar field
$\phi$. The definition differs from ref.\refs{\erg}\ only in
 that here we include the kinetic term $\half(\partial_\mu\phi)^2$
in $S_{\Lambda^f_0}$ (and will do likewise in $\Gamma_\Lambda$) so that
$S_{\Lambda^f_0}$
is the {\sl full} bare action for the theory, while
 $C^{-1}(q,\Lambda)=(1/\te q -1)\ q^2$
is now an `additive' I.R. cutoff. Here, $\te q$ is a smooth
regularisation of the Heaviside $\theta$ function, of width $2\epsilon$,
satisfying $0<\te q<1$ for all (positive) $\Lambda$ and
$q$,  but with $\te q\to \theta(q-\Lambda)$ as $\epsilon\to0$. The additive
form for the cutoff
 will make manifest the field reparametrization invariance:
that is invariance of the flow equations under rescaling of the field \resca.
 (This fact was already used in ref.\refs{\ui}.
The same effect would be obtained by redefining
$\Gamma_\Lambda\mapsto\Gamma_\Lambda-\half\int\!d^Dx\,(\partial_\mu\phi)^2$
at the end of the calculation.) The reason for this can be
understood as follows: For any finite $\epsilon$ the invariance is broken
by the induced change in the cutoff term
$C^{-1}\mapsto\lambda^2 C^{-1}$,  but
in the limit of sharp cutoff (i.e. $\epsilon\to0$),
$\lambda^2 C^{-1}$ has the same effect
for any  $\lambda$, since it is either zero (for momenta $q>\Lambda$)
or forces the integrand of \zorig\ to vanish.
{}From \zorig\ we have
$${\partial\over\partial\Lambda}W_\Lambda[J]=
-{1\over2}\left\{ {\delta W_\Lambda\over
\delta J}.{\partial C^{-1} \over\partial\Lambda}.{\delta
W_\Lambda\over\delta J} +
\tr\left({\partial C^{-1} \over\partial\Lambda}.{\delta^2
W_\Lambda\over\delta J\delta J}\right)\right\}\quad ,$$
which on rewriting in terms of the Legendre effective action $\Gamma_\Lambda$
gives
(as in ref.\refs{\erg}),
\eqn\begn{{\partial\over\partial\Lambda}\Gamma_\Lambda[\phi]=
{1\over2}\int\!{d^Dq\over (2\pi)^D}\
{\partial C^{-1}(q,\Lambda)\over \partial\Lambda}\left[C^{-1}+
{\delta^2\Gamma_\Lambda\over\delta\phi\delta\phi}\right]^{-1}
\mkern-23mu(\q,-\q)\quad.}
$\Gamma_\Lambda$ is defined by
$\Gamma_\Lambda[\phi]+\half\phi.C^{-1}.\phi=-W_\Lambda[J]+J.\phi$,
where now $\phi=\delta W_\Lambda/\delta J$ is the classical field.
In the limit $\epsilon\to0$, the $\partial C^{-1}(q,\Lambda)/\partial\Lambda$
term restricts the momentum integral to the spherical shell $q=\Lambda$.
Therefore any other terms in \begn\ containing $C^{-1}(q,\Lambda)$ [or
$C^{-1}(\Lambda,\Lambda)$], become ambiguous in this limit, since they in turn
contain $\theta(0)$, which is ill-defined.\foot{$\theta(0)$ is by no means
simply given by $\theta(0)=\half$ \refs{\erg} !}\ To properly resolve the sharp
cutoff limit we must therefore isolate all such terms and treat these more
carefully\refs{\erg}. Thus, following ref.\refs{\erg},
we separate from the two-point
function the field independent full inverse propagator $\gamma(p,\Lambda)$:
\eqn\spl{{\delta^2\Gamma_\Lambda[\phi]\over\delta\phi\delta\phi}(\p,\p')
=\gamma(p,\Lambda) (2\pi)^D\delta(\p+\p')+{\hat\Gamma}[\phi](\p,\p'; \Lambda)
\quad,}
so that ${\hat\Gamma}[0]=0$, and drop from both sides of \begn\
the field independent vacuum energy contribution. The subtracted form of \begn\
can then be written
\eqn\middl{{\partial\over\partial\Lambda}\Gamma_\Lambda
=-{1\over2}\tr\left\{{\partial C^{-1}\over \partial\Lambda}
(C^{-1}+\gamma)^{-2}.
{\hat\Gamma}.\left(1+[C^{-1}+\gamma]^{-1}\!.
{\hat\Gamma}\right)^{-1}\right\}\quad.}
We will assume that the square-bracketed $[C^{-1}+\gamma]$ term, buffered as it
is on both sides by ${\hat\Gamma}[\phi]$, `almost never' carries a
momentum $p=\Lambda$, i.e. that such points form a set of zero measure for the
integrations in \middl. (The validity of this `zero measure assumption'
 will be addressed in
sect.5.) Then, from the definition of $C^{-1}(p,\Lambda)$ we
may safely take the limit
\eqn\gr{\lim_{\epsilon\to0}{1\over C^{-1}(p,\Lambda)+\gamma(p,\Lambda)}=
G(p,\Lambda)\quad,}
where $G(p,\Lambda)=\theta(p-\Lambda)/\gamma(p,\Lambda)$ is the I.R. cutoff
full
Green function. To complete the sharp cutoff limit of \middl\ we need to note
$$\eqalign{\lim_{\epsilon\to0}{\partial C^{-1}(q,\Lambda)\over\partial\Lambda}
\left[C^{-1}(q,\Lambda)+\gamma(q,\Lambda)\right]^{-2}
&=-{\partial\over\partial\Lambda}\left\{\lim_{\epsilon\to0}{1\over
C^{-1}(q,\Lambda)+\gamma(q,{\tilde\Lambda})}\right\}
\Big|_{{\tilde\Lambda}=\Lambda}\cr
&=\delta(q-\Lambda)/\gamma(q,\Lambda)\quad,}$$
where the last line results from taking the limit in a similar way to \gr.
Substituting this and \gr\ in \middl\ we obtain the sharp cutoff flow equation:
\eqn\alm{{\partial\over\partial\Lambda}\Gamma_\Lambda
=-{1\over2} \int\!{d^Dq\over (2\pi)^D}\ {\delta(q-\Lambda)\over
\gamma(q,\Lambda)} \left[{\hat\Gamma}.
(1+G.{\hat\Gamma})^{-1}\right]\!(\q,-\q)\quad.}
Under the field rescaling \resca\ we have that ${\hat
\Gamma}\mapsto\lambda^2{\hat\Gamma}$,
$\gamma\mapsto\lambda^2\gamma$ and $G\mapsto G/\lambda^2$, as trivially
follows from \spl.
We see that the sharp cutoff flow equation \alm\ indeed has manifest invariance
under field rescaling \resca, as promised.
 This equation may now be expanded as a power series in the field $\phi$ to
yield an infinite set of coupled non-linear flow equations for the ($n$-point)
one particle irreducible Green functions $\Gamma(\p_1,\cdots,\p_n;\Lambda)$.
As usual, in momentum space we factor out the momentum conserving
$\delta$-function   [$\Gamma(\p,-\p;\Lambda)\equiv\gamma(p,\Lambda)$]:
\eqn\onepi{(2\pi)^D\delta(\p_1+\cdots+\p_n)\,\Gamma(\p_1,\cdots,\p_n;\Lambda)=
{\delta^n\Gamma_\Lambda[\phi]\over\delta\phi(\p_1)\cdots\delta\phi(\p_n)}
\quad.}

Our interest in using these equations is to investigate continuum limits, which
are found by approaching a fixed point as $\Lambda\to0$. This justifies
dropping all reference to the overall cutoff $\Lambda^f_0$, which we do
implicitly (explicitly in ref.\refs{\erg}), since all momenta
can be restricted to be very much less than $\Lambda^f_0$, as is clear from
the trivial boundedness of the momentum integral in \alm.
At the fixed point there is only one mass scale, namely $\Lambda$, and
therefore
it is helpful to write all dimensionful quantities in terms of this. We need
only observe that the scaling dimension $d_\phi$ of the field $\phi$, at the
fixed point, is
generally anomalous: $d_\phi=\half(D-2+\eta)$, where
 $\eta$ is the anomalous scaling
dimension. We could scale $\Lambda$ out of the infinite set of flow equations
for the $\Gamma(\p_1,\cdots,\p_n;\Lambda)$, deducing their scaling dimensions
from \onepi, but it is more elegant
to scale $\Lambda$ directly out of the functional
eqn.\alm. Therefore, similarly to refs.\refs{\deriv,\trunc},
we write $\q\mapsto
\Lambda\q$, $\phi(\Lambda\q)\mapsto \Lambda^{d_\phi-D}\phi(\q)/\zeta$,
$\Gamma_\Lambda[\zeta^{-1}\Lambda^{d_\phi}\phi(\x)]\mapsto\zeta^{-2}\Gamma_t[
\phi]$, $\gamma(\Lambda p,\Lambda)\mapsto\Lambda^{2-\eta}\gamma(p,t)$,
${\hat\Gamma}[\zeta^{-1}\Lambda^{d_\phi}\phi](\Lambda\p,\Lambda\p';\Lambda)
\mapsto \Lambda^{2-D-\eta}{\hat\Gamma}[\phi](\p,\p';t)$ and
$t=\ln(\Lambda_0/\Lambda)$. Here $\Lambda_0$ is the energy scale
at which the `bare' $\Gamma_0[\phi]$ will be defined
($\Lambda_0<\!\!<\Lambda_0^f$), and the
number $\zeta=(4\pi)^{D/4}\sqrt{\Gamma(D/2)}$ is chosen for convenience.
(The physical meaning of $\Gamma_0[\phi]$ is discussed in sect.4). In
addition we perform the radial part of the $q$ integral in \alm. The result is:
\eqn\fnl{({\partial\over\partial t}+d_\phi\Delta_\phi+\Delta_p)\,\Gamma_t[\phi]
={1\over\gamma(1,t)}\left\langle\left[{\hat\Gamma}.
(1+G.{\hat\Gamma})^{-1}\right]\!(\q,-\q)\right\rangle\quad.}
In here, $\Delta_\phi=\phi.{\delta\over\delta\phi}$ is the `phi-ness' counting
operator: it counts the number of occurences of the field $\phi$ in a
given vertex. $\Delta_p$ may be expressed as
\eqn\momsc{\Delta_p =\int\!{d^Dp\over(2\pi)^D}\,\phi(\p)p^\mu
{\partial\over\partial p^\mu} {\delta\over\delta\phi(\p)}}
and is the momentum scale counting operator. The angular brackets stand for an
average (i.e. normalised angular integration) over all directions of the unit
vector $q=1$. The full propagator $G(p,t)$ is now given by
\eqn\prop{G(p,t)=\theta(p-1)/\gamma(p,t)\quad.}
Using $(1+G.{\hat\Gamma})^{-1}=1-G.{\hat\Gamma}+(G.{\hat\Gamma})^2-\cdots$
in \fnl\ and expanding  in $\phi$ we obtain the infinite set of
scaled flow equations for the $n$-point Green functions:
\eqn\gflow{\eqalign{&\left({\partial\over\partial t}
+\sum_{i=1}^np_i^\mu{\partial\over\partial p_i^\mu}+nd_\phi-D\right)
\Gamma(\p_1,\cdots,\p_n;t)={1\over\gamma(1,t)}
\big\langle\Gamma(\q,-\q,\p_1,\cdots,\p_n;t)\big\rangle\cr
&-{2\over\gamma(1,t)}\sum_{\{I_1,I_2\}}\big\langle
\Gamma(\q,-\q-\P_1,I_1;t)\,G(|\q+\P_1|,t)\,
\Gamma(\q-\P_2,-\q,I_2;t)\big\rangle\cr
&+{2\over\gamma(1,t)}\sum_{\{I_1,I_2\},I_3}\big\langle
\Gamma(\q,-\q-\P_1,I_1;t)\,G(|\q+\P_1|,t)\times\cr
&\hskip 1cm\Gamma(\q+\P_1,-\q+\P_2,I_3;t)\, G(|\q-\P_2|,t)\,
\Gamma(\q-\P_2,-\q,I_2;t)\big\rangle+\cdots\quad.}}
These are
illustrated in \fig\oneflo{The sharp cutoff flow equations for
one particle irreducible vertices. Internal lines are full
propagators. The black dot now represents restriction to momentum $q=1$;
the other  propagators have an I.R. momentum cutoff $p>1$.
The symbol $\Delta$ represents the operator on the LHS (Left Hand Side) of
eqn.\gflow}.  $\P_i=\sum_{\p_k\in I_i}\p_k$, and
$\sum_{\{I_1,I_2\},I_3,\cdots,I_m}$ is
a sum over all disjoint subsets $I_i\cap I_j=\emptyset$ ($\forall i,j$)
such that
$\bigcup_{i=1}^m I_i=\{\p_1,\cdots,\p_n\}$. The symmetrization $\{I_1,I_2\}$
means this pair is counted only once i.e. $\{I_1,I_2\}\equiv\{I_2,I_1\}$.
Evidently the expansion stops at the term where all vertices have their
minimum number of legs, i.e. at the $n^{\rm th}$ term in general, or the
$(n/2)^{\rm th}$ term in the $\phi\leftrightarrow-\phi$ invariant theory.
Their structure is identical to that given in ref.\erg,
except for scaling out of $\Lambda$
and the redefinitions to exhibit field reparametrization invariance (not
discussed in \erg).
The latter now appears, by \onepi\prop,  simply as `$n$-point' invariance:
\eqn\npoint{\Gamma(\p_1,\cdots,\p_n;t)
\mapsto\lambda^n \Gamma(\p_1,\cdots,\p_n;t)\hskip2cm n\ge2\quad.}
Of course \fnl\ and \gflow\ are still exact, however these are
now in a particularly helpful form for considering the momentum expansion.

\bigskip\medskip
\centerline{
\epsfxsize=\hsize\epsfbox{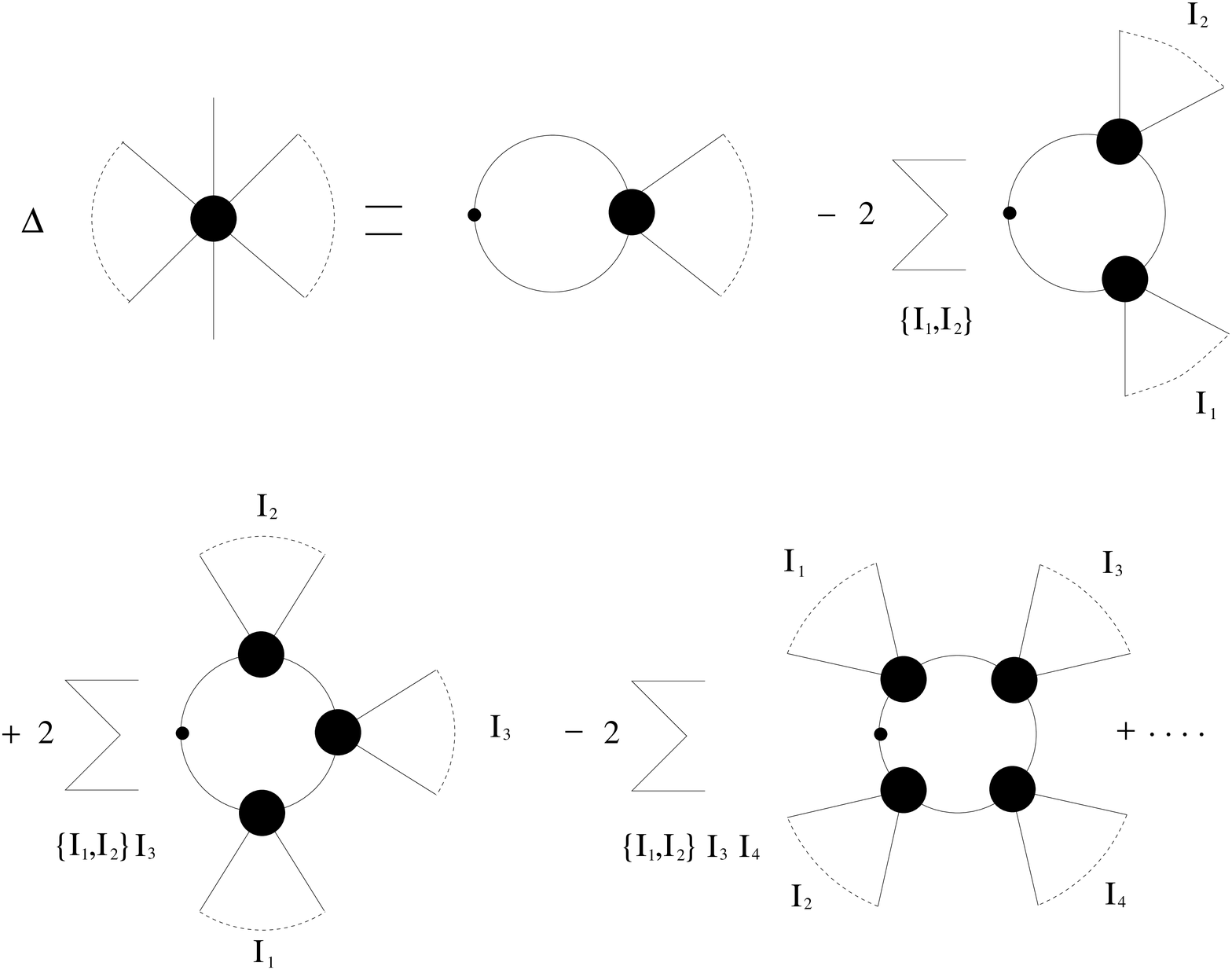}}  
\bigskip
\centerline{\vbox{\noindent {\bf Fig.2.}
The sharp cutoff flow equations for
one particle irreducible vertices. Internal lines are full
propagators. The black dot now represents restriction to momentum $q=1$;
the other  propagators have an I.R. momentum cutoff so that $p>1$.
The symbol $\Delta$ represents the operator on the LHS (Left Hand Side) of
eqn.\gflow.
}}
\bigskip\medskip

\newsec{The momentum scale expansion.}
The momentum expansion
for the Green functions $\Gamma(\p_1,\cdots,\p_n;t)$ is in terms of homogeneous
functions of non-negative integer degree\erg:
\eqn\momexp{\eqalign{\Gamma(\p_1,\cdots,\p_n;t) &=\sum_{m=0}^\infty
\Gamma^{(m)}(\p_1,\cdots,\p_n;t)\hskip2cm\cr
\ins{.5}{.5}{such that}\Gamma^{(m)}(\rho\p_1,\cdots,\rho\p_n;t) &=\rho^m
\Gamma^{(m)}(\p_1,\cdots,\p_n;t)\quad.}}
Or, before expanding in $\phi$, we may write equivalently:
\eqn\momexpi{
\Gamma_t[\phi] =\sum_{m=0}^\infty \Gamma_t^{(m)}[\phi]
\ins11{where}\Delta_p\Gamma_t^{(m)}[\phi] =(m-D)\Gamma_t^{(m)}[\phi]}
[hence the  name ``momentum scale counting operator'' \momsc.
The extra factor of $D$ arises from the momentum conserving $\delta$-function
in \onepi.] All other external momentum dependence in \gflow\
may be expanded as a  series in integer degree homogeneous functions
by introducing the ``momentum scale'' expansion
 parameter $\rho$, and expanding as a power series in $\rho$,
e.g. the one particle irreducible Green functions generally are reexpanded
in $\rho$ via $\Gamma^{(m)}(\q+\rho\P,-\q-\rho\P',\cdots)$,
after which $\rho$ may be reset to one. The I.R. cutoffs thus have
expansions as:
\eqn\expIR{
\theta(|\P+\q|-1)=\theta(\q.{\hat\P}+P/2)=
\theta(\q.{\hat\P})+\sum_{m=1}^\infty{1\over
m!}\,({P/2})^m\,\delta^{(m-1)}(\q.{\hat\P})\quad,}
where we have introduced the unit vector ${\hat\P}=\P/P$ (and used
$\theta(x-1)\equiv\theta(x^2-1)$. $\delta^{(m)}(x)$ means the $m^{\rm th}$
derivative of the $\delta$-function with respect to $x$.)
Of course the fact that {\sl all} external momentum dependence in \gflow\
can be expanded in this way
in terms of homogeneous
functions of non-negative integer degree, guarantees that
such an integral momentum scale expansion is
self-consistent
(as opposed to needing for example  homogeneous
functions of fractional degree).
A systematic sequence of approximations -- the $O(p^M)$
approximations -- now results from replacing the sum $\sum^\infty_{m=0}$
in \momexp\momexpi\
by $\sum_{m=0}^M$, a sum up to some maximum power of momentum scale
$M=0,1,2,\cdots$\ , substituting this into the flow equations \fnl\gflow,
and expanding the RHS (Right Hand Side) of these equations
up to the same order in momentum scale.

If the $\Gamma(\p_1,\cdots,\p_n;t)$ were analytic functions
of momenta at the origin $\p_i={\bf 0}$, then they would have a Taylor
expansion in the momenta, equivalent to a derivative expansion in position
space, and therefore odd $m$ $\Gamma^{(m)}$ would vanish and
even $m$ $\Gamma^{(m)}$ would be sums of Lorentz invariant products of $m$
momenta.
Such an expansion is not however possible for sharp cutoff\refs{\kogwil,\erg},
a point we reemphasise here. The sharp cutoff induces non-analyticity at the
origin 
of momentum space. This is already evident in
the expansion \expIR: the odd $m$ terms do not vanish and the $O(p^0)$'th
term, $\theta(\q.{\hat\P})$, is still a function of the angle between between
$\P$ and $\q$ and not simply a constant as would be required by analyticity.
Remarkably, this non-analytic dependence cancels out at $O(p^0)$, as we will
see in sect.5.
For an expansion in small momenta {\sl beyond} $O(p^0)$, there is
no such serendipity, and the best that one can arrange is an expansion in
momentum scale of integer degree as in \momexp\momexpi. This can be readily
confirmed by computing the momentum expansion beyond lowest order, on
perturbatively evaluated Green functions (with I.R.
cutoff internal propagators) e.g. the 4-point function to one loop in
$\lambda\phi^4$ theory (as was done in effect in ref.\refs{\erg}
and reproduced here in appendix A, for completeness).
At this point we should also recall that this non-analyticity
is only a technical problem, and does not indicate that there
is anything fundamentally wrong with a sharp cutoff\refs{\erg}.

\newsec{Uniqueness and locality.}
So far we have indicated how to form the $O(p^M)$ approximations to the
flow equations \gflow. From their structure as first order differential
equations in $t$, it is clear that they serve to determine {\sl uniquely}
the $O(p^M)$
approximation to the Green functions $\Gamma^{(m)}$, $m=0,\cdots,M$,
at `time' $t+\delta t$, from their values at time $t$, and hence uniquely
from  the
`initial' values of the $\Gamma^{(m)}$ at time $t=0$
(i.e. $\Lambda=\Lambda_0$).
It is here (without loss of generality) that we must recover the notion of
locality. The point is that we want the full theory
\zorig, in the limit $\Lambda\to0$, not to suffer from these spurious
non-analyticities at zero momenta (which correspond to spurious non-local
behaviour in position space). In the continuum limit,
this behaviour should be purely a consequence
of inserting a sharp low energy
I.R. cutoff in the path integral, and cancel out once
the momentum modes with $p<\Lambda$ are also included. Locality would
usually be implemented by insisting that the bare action
$S_{\Lambda^f_0}$ be local (that is, be expressible in terms of
 a derivative expansion). However this would require reintroducing the overall
cutoff $\Lambda_0^f$, complicating the flow equations
somewhat\refs{\erg}, and this we wish to avoid. Therefore we will impose
instead, at $\Lambda=\Lambda_0$, that the `bare' Legendre action
$\Gamma_0[\phi]$ be local\refs{\erg}. By universality (of the continuum limits)
we do not expect that this change matters for renormalized
quantities, since it amounts to some minor alterations in how the U.V.
cutoff is imposed. [Feynman diagrams vanish if {\sl any} 
internal momentum is larger than $\Lambda_0^f$, while otherwise
these diagrams vanish only if
{\sl all} internal momenta are larger than  $\Lambda_0$ \refs{\erg}.
The reason for this difference is simple: the latter is the complement to the
requirement that is imposed by the infrared cutoff when $\Lambda=\Lambda_0$
namely that the contribution would be non-vanishing only if all 
internal momenta are greater than $\Lambda_0$.
That the weaker U.V. constraint at $\Lambda_0$
already fully regularises the theory, is clearest from
the boundedness of the momentum integrals in \alm\fnl\gflow.]
Using \gflow\ to evolve away from $t=0$, non-analyticity will be generated
by the terms in \expIR, however this non-analyticity is now fixed,
 up to the usual arbitrariness in the choice of local bare action, and
acceptable in the sense that it can be expected to disappear as
$\Lambda/\mu\to0$ (where $\mu<\!\!<\Lambda_0$ is
some typical scale of low energy physics).
This is a full `in principle' solution to fixing the momentum dependence
of the $\Gamma^{(m)}$, but, providing the angular averages on the RHS of
\gflow\
can be performed, we find in practice that much more can be done, as follows.

One method is simply to ansatz the general momentum dependence of the
$\Gamma^{(m)}$ (with appropriate $t$ dependent coefficients). If the ansatz
solves
the equations \gflow, in the sense that the momentum dependence of both
sides of the equation can be made to agree (within the $O(p^M)$ approximation)
by appropriate choice of the flow equations for the coefficients, and if the
ansatz is sufficiently general that it matches the local bare action
[up to $O(p^M)$] at $t=0$ (by appropriate choice of boundary conditions for
the coefficients\foot{Of course any terms which are then zero
for all $t$,  are dropped.}),
then it is {\sl the} solution, by the uniqueness properties of the first
order in $t$ differential equation (as discussed above).

Another method is to
determine the solution systematically as follows.
We have that at $t=0$ the $\Gamma^{(m)}$ are non-vanishing only for $m$ even,
 and the latter are generally arbitrary sums of products of $m$ momenta
[subject only to the general constraints of Lorentz invariance, permutation
invariance, momentum conservation etc. as follows from e.g. \onepi].
Substituting these  general expressions for $\Gamma=\sum_{m=0}^M \Gamma^{(m)}$
into \gflow, and
performing the momentum scale expansion, we can determine the general form
of the non-analytic parts in $\Gamma^{(m)}$ at some small time later
$t=\delta t$. From this we can deduce at the `linearised level' the general
$O(p^M)$ form of $\Gamma$ at any time $t$ by replacing those coefficients of
order $\delta t$ by general coefficients. Now we substitute {\sl this}
expression into the RHS of \gflow\ and hence deduce the general form of the
$O(p^M)$ $\Gamma$ at the `quadratic level'. Iterating, we find in practice
that this procedure converges and in this way
we deduce the general $O(p^M)$ form of
the $\Gamma^{(m)}$ as a linear superposition of a finite set of `basis'
functions
of momenta. Only the coefficients carry the $t$ dependence and of course
those multiplying non-analytic momentum functions, must vanish at $t=0$.
 In the next two
sections we give examples of $O(p^M)$ approximations.

\newsec{The local potential approximation.} 
We will now work out the $O(p^0)$ approximation. Since truncations of the
field dependence tend to have limited accuracy and
reliability\refs{\trunc\revi}, we make no further approximation.

It is here that we must discuss the validity of the zero measure assumption,
made below
\middl, which is equivalent to assuming  that the ill-defined
$p=1$ propagator $G(1,t)=\theta(0)/\gamma(1,t)$ [from \prop]
appears only at a set of points of zero measure in \fnl. [{\sl If this is
the case} then
the result is the same whatever finite value one assigns to these
$\theta(0)$'s, after performing the integrals in \fnl.]  Actually this
assumption is plainly wrong when $\phi(\x)$ obtains
a non-zero spatially independent vacuum expectation value $\phi(\x)=\vev{\phi}$
\refs{\erg}, since then ${\hat\Gamma}[\phi]$ has a non-zero `diagonal' part
i.e. a part
proportional to $\delta(\p+\p')$. In ref.\refs{\erg}\ we handled this by
redefining $\gamma$ in \spl\ to absorb the diagonal part. One of the main
results of this paper however will be to show that this procedure is not
necessary. Instead, we will show that at $O(p^0)$, the vanishing external
momentum limit
of the sharp cutoff equations (i.e. considering $\phi(\x)\to
\vev{\phi}$, {\sl after} having taken $\epsilon\to0$), is unique and equal to
the result of taking the $\epsilon\to0$ limit of the equations
where all momentum dependence is discarded first (as was done in
refs.\refs{\erg,\trunc,\ui}).
In other words we will show that these two limits commute. As we will see, this
fact is by no means trivial however: the agreement is obtained only at the
end of the computation and in very different ways in the two methods.
What is more, the $O(p^0)$ approximation exactly
coincides with the local potential
approximation\refs{\nico} \LPA,
providing further justification for the ubiquitous
use of the latter\tons\ by demonstrating that it is simply the start of the
systematic sequence of $O(p^M)$ approximations. Where derivations of the  local
potential approximation appear
in refs.\tons, they are all of the same type as that in
refs.\refs{\wegho,\nico,\hashas}:
utilising a finite width momentum shell (with sharp edges) and discarding
the momentum dependence first, before allowing the width to tend to zero.
 This is equivalent in effect
to the derivation in refs.\refs{\erg,\trunc}\ if  one specializes to a
cutoff function that simply linearly interpolates between
constant values outside the shell i.e. to
$\te q=(q-\Lambda)/\epsilon$ for $|q-\Lambda|<\epsilon$.
The important point is that in all these derivations it is not possible
even in principle to improve
on the approximation of discarding all the momentum dependence, while the
latter
is done before letting the width $\epsilon\to0$. This is simply
because any non-zero
external momentum $p$ will satisfy $p>\!\!>\epsilon$ as $\epsilon\to0$,
but proper
consideration of this regime precisely corresponds to interchanging the
limits. Therefore any attempt to go beyond the simplest approximation
of discarding all momentum dependence, must first
address the question of whether
this exchange of limits still yields the same answer for the simplest
approximation. As we have already stated, this turns out to be non-trivial
but true.

At $O(p^0)$ the {\sl local bare} action can only include interactions in a
bare potential $V_0(\phi)$ which however is taken to be completely general
to begin with.\foot{Higher terms in momentum scale can be added, but
at $O(p^0)$ there is
no feedback into these terms from the RHS of the flow equations \fnl\gflow.
The LHS of these equations then constrain these terms to be singular in $\phi$
(which is not allowed)
or vanish. See ref.\deriv\ or ref.\twod\ for further explanation.}\
  By either of the methods discussed in sect.4,
this suggests that we start by ansatzing
\eqn\opo{\Gamma_t[\phi]=\int\!d^D\!x\,\left\{\half
 (\partial_\mu\phi)^2+V_t(\phi)\right\}\quad.}
This will turn out to be the complete solution to the momentum dependence
at $O(p^0)$, i.e. the vertices $\Gamma(\p_1,\cdots,\p_n;t)$ with $n>2$,
which are in principle still
functions of the angles between the external momenta,  may be taken to be
simply constants. Note that the kinetic term
will not get corrected at $O(p^0)$ [or $O(p^1)$] so carries no $t$ dependence.
The normalization for the kinetic term is conventional, but would require
(and could {\sl at this order} be given) separate justification here
 if field reparametrization invariance had been broken,
since then the normalization in general affects the
results\refs{\oldrepar,\gol,\deriv,\revi}.
However in this case we have immediately that
 all normalizations are equivalent, by the invariance \resca.
Equation \opo\ implies that
\eqnn\gert
\eqnn\freda
$$\eqalignno{\gamma(p,t)&=p^2+V''_t(0) &\gert\cr
\ins{0}{1}{and}{\hat\Gamma}[\phi](\p,-\p-\P;t)\equiv {\hat\Gamma}(\P)
&= \int\!d^D\!x\, \left[V''_t\!\left(\phi(\x)\right)
-V''_t(0)\right]\,\e{i\P.\x}\quad. &\freda\cr}$$
Here, prime refers to partial differentiation with respect to
$\phi$. The latter formula follows directly from
$\delta^2\Gamma/\delta\phi(\p)\delta\phi(\p')$ $\equiv$
$\int\! d^D\!xd^D\!y\,\e{-i(\p.\x+\p'.\y)}
\delta^2\Gamma/\delta\phi(\x)\delta\phi(\y)$.
We have introduced the notation ${\hat\Gamma}(\P)$ to emphasise
that ${\hat\Gamma}$ depends only on the sum of its two momentum arguments;
this is a consequence of dropping all momentum dependence in the
vertices.
Substituting this, \opo\ and \prop\ into \fnl, expanding the
RHS of \fnl\ via
$(1+G.{\hat\Gamma})^{-1}=1-G.{\hat\Gamma}+(G.{\hat\Gamma})^2-\cdots$, and
dropping all but the $O(p^0)$ terms in $\gamma$ and $\theta$
in the RHS (c.f. \momexpi,\expIR\
and the discussion inbetween) one obtains:
$${\eta\over2}\int\!d^D\!\x\,  (\partial_\mu\phi)^2=0\quad,$$
implying $\eta=0$ for this approximation, as expected, and:
\eqnn\bert
$$\eqalignno{&\int\!d^D\!x\,\left\{{\partial\over\partial t}
V_t(\phi)+{1\over2}(D-2)\phi V'_t(\phi)-DV_t(\phi)\right\}=&\bert\cr
&{{\hat\Gamma}({\bf 0})\over\gamma(1,t)}
   -\sum_{r=2}^\infty
{(-1)^r\over\gamma(1,t)^r}\int{d^D\!P_1d^D\!P_2\cdots d^D\!P_{r-1}
\over (2\pi)^{D(r-1)}}\,{\hat\Gamma}(\P_1){\hat\Gamma}(\P_2)\cdots
{\hat\Gamma}(\P_r)\cr
&\hskip 3.5cm\times\big\langle\theta\!\left(\q.\P_1\right)
\theta\!\left(\q.[\P_1+\P_2]\right)\cdots
\theta\!\left(\q.[\P_1+\P_2+\cdots+\P_{r-1}]\right)\big\rangle\quad,}$$
where $\P_r=-\sum_{k=1}^{r-1}\P_k$. The $r^{\rm th}$ contribution can
be represented diagrammatically as in
\fig\pzero{cf p(xi) of notes for paper; Recall that the small dot indicates
averaging over all directions of the propagators momentum which is
resticted to be unity}, and alternatively
follows from the $O(p^0)$
parts of contributions in \oneflo, after integration over the fields
$\phi(\p_i)$ and resummation over $n$.
\midinsert
\centerline{
\epsfxsize=0.6\hsize\epsfbox{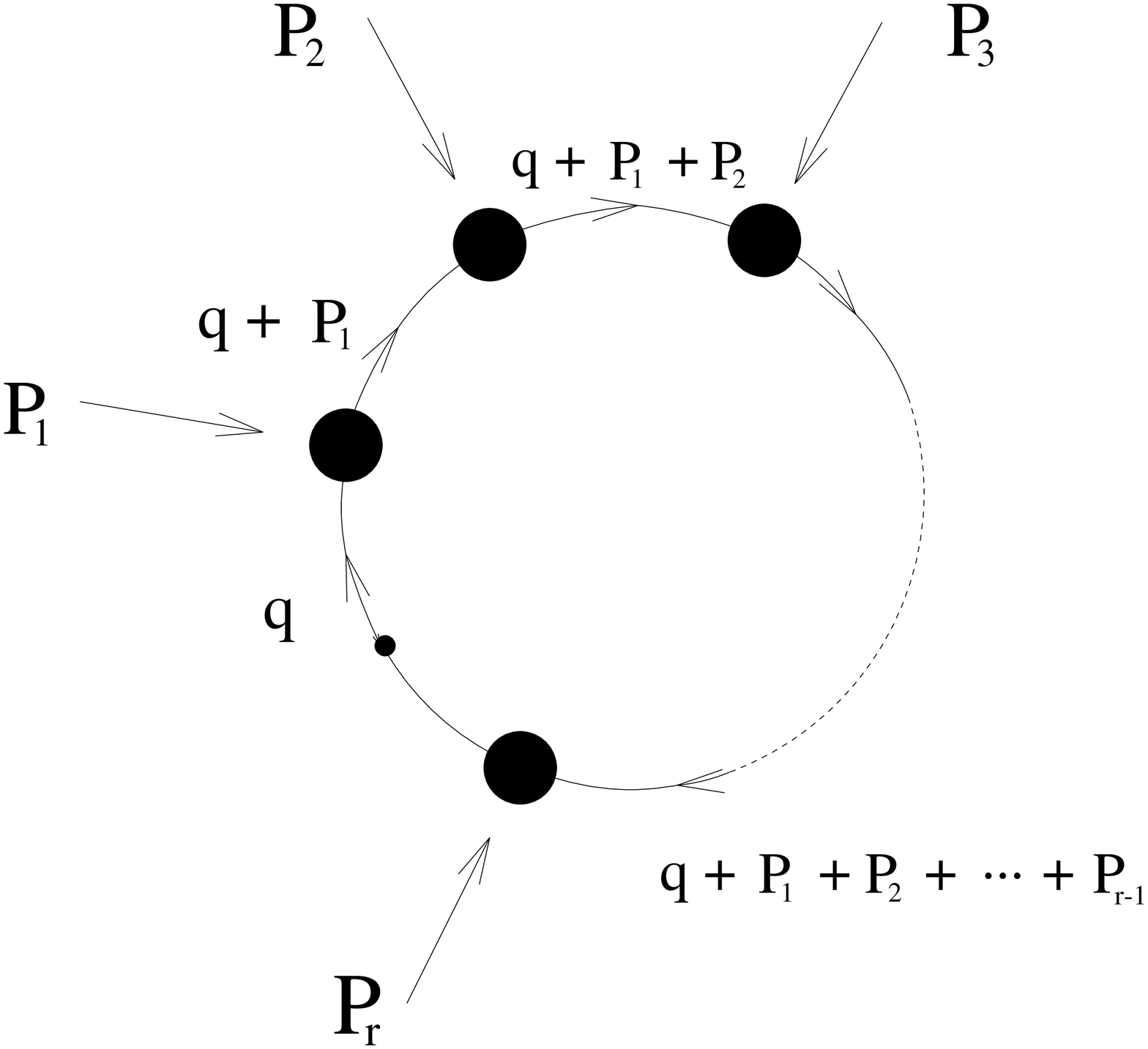}}  
\bigskip
\centerline{\vbox{\noindent {\bf Fig.3.}
The $r^{\rm th}$ contribution to eq.\bert, represented diagrammatically.
 Recall that the small dot indicates
averaging over all directions of the propagators momentum $\q$, which is
resticted to be of unit norm.
}}
\endinsert
The above average over the
unit vector $\q$ is indeed a
non-trivial function of the angles between the momenta $\P_k$, but by
utilising the cyclic symmetry of \pzero\ or \bert\ with respect to the external
momenta, we can effectively replace its contribution by $1/r$, as follows.
We rewrite the RHS of \bert\ as:
\eqn\Qit{
-\sum_{r=1}^\infty(-1)^r\int\left(\prod_{k=1}^r {d^D\!P_k\over (2\pi)^D}
{{\hat\Gamma}(\P_k)\over\gamma(1,t)}\right) (2\pi)^D
\delta\!\left(\sum_{k=1}^r\P_r\right)
\big\langle\theta\!\left(\q.\Q_2\right)
\theta\!\left(\q.\Q_3\right)\cdots
\theta\!\left(\q.\Q_r\right)\big\rangle\quad,}
where we have introduced the partial sums $\Q_k=\sum_{j=1}^{k-1}\P_j$ for
$2\le k\le r$, $\Q_1={\bf 0}$, and defined
the angular average to be unity for $r=1$.
Taking the average of the $r$ integrals obtained by cyclically
relabelling the integration variables $\P_1,\cdots,\P_r$ (geometrically
corresponding to the $r$ equivalent diagrams obtained from
\pzero\ by placing the small dot on each of the $r$ propagators) we obtain:
\eqn\alfie{-\sum_{r=1}^\infty {(-1)^r\over r}\int
\left(\prod_{k=1}^r {d^D\!P_k\over (2\pi)^D}
{{\hat\Gamma}(\P_k)\over\gamma(1,t)}\right) (2\pi)^D
\delta\!\left(\sum_{k=1}^r\P_r\right) \left\langle\sum_{j=1}^r
\prod_{{k=1}\atop{k\ne j}}^r\theta\!\left(\q.[\Q_k-\Q_j]\right)
\right\rangle\ .}
Now, generically for any given vector $\q$, one of the $\q.\Q_k$,
say $\q.\Q_i$, will be less than all the others i.e.
$\q.\Q_i<\q.\Q_k\quad$ $\forall k\ne i \in(1,\cdots,r)$. It follows that for
this $\q$,
the $j=i$ contribution to the above $j$-sum is unity, and the $j\ne i$
contributions vanish. Therefore,
except for a set of points of zero measure for the $q=1$
angular average (defined by those vectors $\q$ such that
$\q.\Q_i=\q.\Q_k$ for some $k$),
$\sum_{j=1}^r
\prod_{k\ne j}^r\theta\!\left(\q.[\Q_k-\Q_j]\right)=1$. Substituting this,
 and \freda, and
writing
$(2\pi)^D
\delta\!\left(\sum_{k=1}^r\P_r\right)=\int\! d^D\!x\,\exp-i(\x.\P_1+\cdots+
\x.\P_r)$, we have immediately that \alfie\ equals
$-\int\!d^D\!x\, \sum_{r=1}^\infty {1\over r}
\left[V''_t(0)-V''_t(\phi)\right]^r/\gamma(1,t)^r$.
Using \gert, this may be resummed to $\int\!d^D\!x\,\left\{
 \ln\big[1+V''_t(\phi)\big]-\ln\gamma(1,t)\right\}$.
Here, the second term, which corresponds to a vacuum energy,
may be renormalised away by
redefining the field independent part of the potential as
$V_t\mapsto V_t+\E(t)$, where, by \bert, we require
 ${\partial\E/\partial t}
-D\E=-\ln\gamma(1,t)$. It is neater to notice however, that a $\gamma$
dependent term was discarded in \middl\ as a result of
dropping from \begn\ the field independent
vacuum energy contribution. If this $\gamma$ dependent term is kept, it is
straightforward to show that it precisely cancels the
$-\int\!d^D\!x\,\ln\gamma(1,t)$ above. Hence finally,
we have from \bert:
\eqn\eqlpa{{\partial\over\partial t}
V_t(\phi)+{1\over2}(D-2)\,\phi V'_t(\phi)-DV_t(\phi)=
\ln\big[1+V''_t(\phi)\big]\quad.}
This coincides with the local potential approximation\tons\ to the
Wegner-Houghton equations\refs{\wegho}.

\newsec{Beyond lowest order.}

We now proceed to the $O(p^1)$ approximation. Following the iterative method
described in sect.4 we try to proceed by initially ansatzing a
momentum independent form for $\Gamma_t$ as in \opo, \gert\ and \freda.
We then substitute these expressions into \fnl\ to see what form the $O(p^1)$
$\Gamma_t$ takes at the `linearised level'. We see from expanding
$\gamma(p,t)$ in \prop\ and using \expIR, that amongst other terms a set
of averages of $\q.\Q_k$ ($k=2,\cdots,r$) of the form
\eqn\theprob{\big\langle\theta\!\left(\q.\Q_2\right)
\cdots
\theta\!\left(\q.\Q_k\right)\!\q.\Q_k\,
\theta\!\left(\q.\Q_{k+1}\right)\cdots
\theta\!\left(\q.\Q_r\right)\big\rangle\quad,}
have to be computed [where we have used the notation introduced in \Qit].
While we were able to replace the analagous expression in \Qit\
by a constant, using (cyclic) symmetry arguments, we have found
the permutation symmetry of the diagrams
\pzero\ to be insufficient, when $r>3$,
to allow a simple replacement for the above expression. Of course the above
expression can be evaluated exactly, in principle, but the result will be
a very complicated $O(p^1)$ function, so that further insight is necessary
to make progress at this level.

For this reason we now restrict the discussion of momentum dependence to
cases where $r\le3$. This may be `justified' by approximating
the flow equations by truncations of
the field dependence\erg\ of the $O(p^{m>0})$ terms 
to six-point functions or less (in a $\phi\leftrightarrow-\phi$ invariant
theory), i.e. discarding $\sim\phi^8$ terms and higher in
non-zero momentum scale pieces. However, we already know that
truncations of the field dependence yield results that
 do not converge (beyond a certain order) and  generally also result in
spurious fixed points\trunc. Nevertheless,
we will discuss these truncations in order
to provide explicit
examples of the general considerations of the previous sections.
As we will see, the results are not impressive. In particular they are
much worse than the results of the momentum expansion with smooth cutoff
in ref.\deriv.
We attribute this to the truncations of the field dependence, because
this was not employed in ref.\deriv,
the sharp cutoff momentum scale expansion results in a rapidly
convergent (numerical) series in a model two-loop calculation
(\erg\ and app.B), and
intuitively one would expect a 
better convergence for sharp cutoff than for
smooth\erg.
Although we cannot at this stage rule out
the possibility that the sharp cutoff momentum scale expansion
is  `at fault' at the non-perturbative level, we suspect that truncations
of the field dependence of the kinetic term $K(\phi)$ in the smooth
case\deriv\ would produce similarly unimpressive results, which would then
strongly support our contention
 that it is also only the truncations that are at fault here.

Because we will be working to  $O(p^m)$ with $m>0$, only
for certain $n$-point functions, we must 
generalise the discussion of
previous sections to the cases where
$m$ can depend on $n$, i.e. where the equations
\gflow\ are expanded to different order in momentum scale depending on the
value of $n$. This generalisation is however straightforward and will be
explained within the examples below.

We concentrate again\deriv\trunc\ on the
critical exponents for the non-perturbative
Wilson fixed point in three dimensions: the results
will provide a simple measure of how good these approximations are, in a
non-perturbative setting. In three dimensions, averages of functions $f$
that depend  on only one component of $\q$, say $\q.{\hat\P}$ as in \expIR,
satisfy:
\eqn\avez{\big\langle f(\q.{\hat\P})\big\rangle={1\over2}\int^1_{-1}\!\!\!\!
dz\, f(z)
\quad.}
It is of course straightforward to give the analogous formula in any other
dimension, four for example\erg, and all averages we consider can
 be reduced to this (or its analogue) by symmetry arguments, in any dimension.

Consider first the simplest possible truncation that will allow non-trivial
momentum dependence: discarding all but the two-point and four-point
Green functions.\foot{This is closely similar to a model computation in
ref.\erg. Were we also to throw away all momentum dependence\trunc\ it would
be a differential version of the ``$S^4$ model''\kogwil.}\ By \gflow\ this
gives \eqna\fl
$$\eqalignno{
\Big({\partial\over\partial t}+p{\partial\over\partial p}
-2+\eta\Big)\gamma(p,t) &=
{1\over\gamma(1,t)} \big\langle\Gamma(\q,-\q,\p,-\p;t)\big\rangle &\fl a\cr
\Big( {\partial\over\partial t}+\sum_{i=1}^4
p_i^\mu{\partial\over\partial p_i^\mu}+1-2\eta\Big)
\Gamma(\p_1,\p_2,\p_3 ,\p_4;t)&= &\fl b\cr
-{2\over\gamma(1,t)} \big\langle
\Gamma(\q,-\r_{12},\p_1,\p_2;t)\,G(r_{12},t)\,\Gamma(\r_{12}, &-\q,\p_3,\p_4;t)
+(2\leftrightarrow3)+(2\leftrightarrow4)\big\rangle\quad.  \cr}$$
where $\r_{ij}=\q+\p_i+\p_j$, and in the second average the last two terms  are
the same as the first term but with indices swopped as indicated. We have
thrown away the six point contribution
$\sim\big\langle\Gamma(\q,-\q,\p_1,\p_2,\p_3,\p_4;t)\big\rangle$ to \fl{b}.
Working to $O(p^1)$ we start again by ansatzing first that the four-point
function is momentum independent. The analogue of \theprob\ is now,
using \avez, simply
$\big\langle\theta\!\left(\q.\P\right)\!\q.\P\big\rangle=P/4$,
where $\P=\p_1+\p_2$, $\p_1+\p_3$ or $\p_1+\p_4$ depending on the term on
the RHS of \fl{b}. Similarly the $O(p^1)$ term in \expIR\ produces these
$P$'s. Therefore at `linearised level' we have that
\eqn\anzf{\Gamma(\p_1,\p_2,\p_3,\p_4;t)=\alpha_0(t)+\alpha_1(t)\big\{
|\p_1+\p_2|+|\p_1+\p_3|+|\p_1+\p_4|\big\}\quad,}
where flow equations for the $\alpha_i$ are to be determined.
Substituting this into eqn.\fl{a}\  results in $p$ dependence of the form
$\big\langle|\q+\p|\big\rangle=1+p^2/3$ [by \avez], therefore a
 quadratic ansatz remains sufficient for $\gamma(p,t)$ at this order.
Substituting \anzf\ back into \fl{b}\ results in contributions e.g.
of the form
\eqn\opoexp{
\Gamma(\q+\p_1+\p_2,-\q,\p_3,\p_4;t)=\alpha_0 +\alpha_1\big\{2+|\p_1+\p_2|
+\q.(\p_1+\p_2)\big\}+O(p^2)\quad,}
 where we have used momentum conservation
and $q=1$, (and similarly for the other such contributions). Therefore
the RHS of \fl{b}\ reproduces the general form of \anzf; our iterative
procedure has closed at the linearised level. To obtain an approximation
for $\eta$ we will work to $O(p^2)$ with $\gamma(p,t)$, writing
\eqn\anzt{\gamma(p,t)=a(t)p^2+\sigma(t)\quad,}
and expanding \fl{a}\ to $O(p^2)$,
but keep to $O(p^1)$ with $\Gamma(\p_1,\cdots,\p_4;t)$. Substituting \anzf\
and \anzt\ into \fl{}\ we obtain: \eqna\ai  
$$\eqalignno{{\partial\sigma\over\partial t}+(\eta-2)\sigma
&={\alpha_0+2\alpha_1\over\sigma+a} &\ai{a}\cr
{\partial a\over\partial t}+\eta a &={2\over3}{\alpha_1\over\sigma+a}&\ai{b}\cr
{\partial\alpha_0\over\partial t}+(2\eta-1)\alpha_0 &=
-3\left({\alpha_0+2\alpha_1\over\sigma+a}\right)^2&\ai{c}\cr
{\partial\alpha_1\over\partial t}+2\eta\alpha_1&=
-{1\over2}{(\alpha_0+2\alpha_1)(\sigma\alpha_0-a\alpha_0+8\sigma\alpha_1
+4a\alpha_1)\over(\sigma+a)^3}\quad.&\ai{d}\cr}$$
By the locality requirements of sect.4 we have that the bare
$\sigma(0),a(0)$ and
$\alpha_0(0)$ are a priori arbitrary, but $\alpha_1(0)=0$. By the field
reparametrization invariance \resca, equivalently $n$-point invariance \npoint,
we can impose the conventional
 wavefunction renormalization condition $a(\infty)=1$.
(The full renormalized quantities are given by their $t=\infty$ values.)

Had field reparametrization invariance been broken (as happens generally under
such approximations\refs{\oldrepar,\gol,\revi}), one would
discover the following problem: the approximate
physical quantities would differ for different values of
$a(\infty)$, even though they do not in the exact formulation [as follows
by using \resca].
This problem only potentially arises when we seek to go beyond the local
potential approximation, and is not yet generally appreciated in the
corresponding literature, largely because $a(\infty)$ (or its analogue)
is set to unity anyway (in effect by the tuning of
bare parameters or the form of the cutoff), even when the
 justification for this specialization is missing.
To date, only two forms of cutoff function are known that allow
a field reparametrization invariance after the momentum
expansion approximation: the sharp cutoff studied
here, and power-law smooth cutoffs\refs{\deriv,\twod,\revi}
 $C(q,\Lambda)\sim (q/\Lambda)^{2\kappa}$.\foot{ It can be shown that for other
choices the field reparametrizations\refs{\red} only leave the
flow equations invariant if they can carry non-trivial momentum dependence,
and this is then broken by momentum expansion to finite order.}\
The problem is particularly
clear if we study the Wilson fixed point\refs{\kogwil}. We assume that
choices can be found for $\sigma(0),a(0)$ and
$\alpha_0(0)$ that lie on the critical
surface. Then as $t\to\infty$, the four quantities $\sigma(t)$,
$a(t)$, $\alpha_0(t)$ and $\alpha_1(t)$ become $t$ independent and eqns.\ai{}\
collapse to four simultaneous algebraic equations. The problem is however,
that with $\eta$, there are {\sl five} quantities to be determined.
Without field reparametrization invariance, this allows $\eta$ to
be determined only
as a function of $a(\infty)$ (say). With field reparametrization
invariance, these equations correspond to non-linear eigenvalue
equations\refs{\wegho} for the field's anomalous dimension
$\eta$,  which therefore (generically) can have only certain discrete values
-- as it should be.

Solving the eqns.\ai{}\ at a fixed point $\sigma(t)\equiv\sigma$,
$a(t)\equiv a$, etc, where they reduce to algebraic
equations, we obtain that $\alpha_0, \alpha_1$ and $\eta$
can be written as polynomials in $\sigma$, which itself satisfies the quintic
$$3\sigma^5+92\sigma^4-154\sigma^3-159\sigma^2+74\sigma+8=0\quad,$$
except for the Gaussian solution $\sigma=\alpha_i=\eta=0$.
By the scaling \resca\npoint, we have set (in all cases) $a\equiv1$.
The quintic has five real solutions $\sigma=-32.2$, $-0.953$,
$-0.0917$, $0.441$ and $2.15$, however only  $\sigma=-0.0917$ gives a
positive $\alpha_0$ (and hence a physically stable potential).
We assume that the domains of attraction of the
other fixed points do not include physically sensible bare actions with
$a(0)>0$, $\alpha_0(0)>0$ and $\alpha_1(0)=0$.\foot{In general however,
there are no reliable ways to reject
the spurious answers resulting from truncation of the field
dependence\refs{\trunc}. We note that in this case, the other
 values of $\sigma$ also result in wildly large values for the other
parameters.}\   Using $\sigma=-.0917$
we obtain $\alpha_0=.103$, $\alpha_1=.0307$ and $\eta=.0225$.
This last value, being universal, can be compared to the best determinations:
$\eta=.032(3)$ or $.038(3)$, where the first derives from perturbation theory
in fixed dimension and the second from $\epsilon$ expansions\refs{\zinn}.
Computing the linearised perturbations of \ai{}\ about this fixed point,
we find a numerical matrix eigenvalue equation\foot{N.B. $a(t)=1+\delta a(t)$
where $\delta a(t)$ is not fixed.}\ whose eigenvalues are
$\lambda=1.88$, $0$, and $-.390\pm.630i$. The most positive eigenvalue
yields
$\nu=1/\lambda=
.532$ by standard arguments\refs{\kogwil},
differing little from the $S^4$ model\refs{\kogwil} ($\nu=.527$), and should
be compared to the best determinations $\nu=.631(2)$ \refs{\zinn}.
The zero eigenvalue corresponds to the reparametrization symmetry \resca\
and is therefore redundant\refs{\red,\deriv}. Finally, the
irrelevant complex eigenvalues
represent the (first) corrections to scaling for this truncation
$\omega=-\lambda=.390\pm.630i$, which should be compared to
the best determinations $\omega=.80(4)$ \refs{\zinn}.
This eigenvalue is the most
affected by the approximation, as should be suspected.
These results
could probably be improved by making the truncation around the minimum of
the fixed point effective potential\refs{\tetwet,\alford}.
 Our point here however, is  to begin with the
simplest possible
non-trivial example of the general method of momentum scale expansions.

We turn next to more sophisticated truncations, 
although we are limited to truncations of the field dependence
in the momentum dependent sector, as we pointed out above.
First let us generalise the above example by avoiding the truncation of the
potential, that is we include all the higher $n$-point functions
$\Gamma(\p_1,\cdots,\p_n;t)$ with $n\ge6$, at $O(p^0)$, with the corresponding
$n\ge6$ flow equations \gflow\ also evaluated to $O(p^0)$.
In this case it is clear that the ansatz that these higher point functions
are independent of momenta, must satisfy these equations, since
the equations differ from those of sect.5 only in the following ways:
$a(t)\ne1$,
$\eta\ne0$, 
 and the four-point function appears in $O(p^0)$ expressions as
$\alpha_0(t)+2\alpha_1(t)$, as follows from \opoexp\ (and its
generalisations). By sect.4, this is then
the unique answer for this truncation, up to the choice of  bare potential
$V_0(\phi)$, and bare kinetic term $a(0)$,
 for the local bare action. The resulting flow equations
are the higher momentum scale
 equations \ai{b,d}\ and the flow for the potential
which may be resummed to:  \eqna\aii\
$${\partial\over\partial t}
V_t(\phi)+{1\over2}(1+\eta)\,\phi V'_t(\phi)-3V_t(\phi)=
\ln\big[a(t)+\alpha_1(t)\phi^2+V''_t(\phi)\big]\quad.\eqno\aii{a}$$
The changes between \eqlpa\ and this equation just follow from the
differences with sect.5, listed above.
At the fixed point we again set $a(t)\equiv a=1$, using \resca. Equations
\ai{a,b,d}\ are unmodified by the inclusion of the $O(p^0)$
$n\ge6$-point functions [\ai{a} being subsumed in \aii{a}], and
may be solved in terms of $\sigma$ as $\alpha_0=(1+\sigma)([\sigma-3]\eta
-2\sigma)$, $\alpha_1={3\over2}(1+\sigma)\eta$ and
$$
\eta=\sigma{9+7\sigma+2\sigma^2\pm\sqrt{3}(1+\sigma)\sqrt{35-8\sigma}\over
6+21\sigma+14\sigma^2+\sigma^3}\quad. \eqno\aii{b}$$
At a fixed point, $V_t(\phi)\equiv V(\phi)$
therefore depends only on the choice of $\sigma$ which
may be regarded as the $\phi=0$ initial condition, together with
the evenness constraint $V'(0)=0$. As explained previously\refs{\trunc} one
finds that all but a discrete set of
solutions of \aii{}\ are singular (at some finite $\phi$) and therefore
unacceptable. We find only two non-singular
solutions, being the trivial Gaussian solution
$V(\phi)=0$, and an approximation to the Wilson fixed point. Quantitative
methods for computing
 the latter are described in ref.\refs{\trunc}. (See also
refs.\refs{\deriv,\twod}. We used the second of the better expansion
methods in \refs{\trunc}.) It is significant that all spurious results
 have disappeared: this is a benefit of not truncating the field
dependence in the potential\refs{\revi}.
We find $\sigma=-.214$ and $\eta=.0660$.
Solving for small perturbations about this fixed point\refs{\trunc}
 we deduce $\nu=.612$ and $\omega=.91$. As expected, this is a significant
improvement over the crude truncation taken previously. These results
are an improvement over those of the pure local potential
approximation\refs{\pare,\hashas,\trunc}: $\eta=0$, $\nu=.690$
and $\omega=.595$, but are not as good as the $O(p^2)$ smooth cutoff
approximation\refs{\deriv}. Again, we attribute this to the effects here of
truncation of the field dependence in the $O(p^1)$ part.

Now we try to improve the results by truncating the $O(p^1)$ part at the
maximum practical value of $n=6$. Here the computation starts to become
quite involved, but once the general form of the $O(p^1)$ dependence
 is determined, it proceeds similarly to the previous example.
Since our primary purpose is to illustrate the techniques
of the momentum scale expansion, we indicate here only how to determine the
momentum dependence, and then present the results.
Since we now wish to include the $O(p^1)$ dependence of the six-point function,
we need to work explicitly with the $n=6$ case of the flow equations \gflow.
We can determine the $O(p^1)$ dependence by iteration, substituting
the previous solutions for the two-point \anzt\ and four-point functions \anzf.
The six-point function's momentum dependence then arises in particular
 from the sum
over all permutations of the legs $\p_1,\cdots,\p_6$ in the graph
\fig\tgr{ .}, that is from the angular average:
\eqn\yucki{\eqalign{\sum_{ {\cal P}(\p_1,\cdots,\p_6)}\!\!\!\!
\big\langle\Gamma( &\q,-\q-\P_1,\p_1,\p_2;t)\, G(|\q+\P_1|,t)\times \cr
&\Gamma(\q+\P_1,-\q+\P_2,\p_5,\p_6;t)\, G(|\q-\P_2|,t)
\Gamma(\q-\P_2,-\q,\p_3,\p_4;t)\big\rangle\quad.\cr}}
Here ${\cal P}(\p_1,\cdots,\p_6)$ is a permutation of the
six momenta $\p_1,\cdots,\p_6$.
$\P_1=\p_1+\p_2$, $\P_2=\p_3+\p_4$ and $\P_3=\p_5+\p_6$.
\midinsert
\centerline{
\epsfxsize=0.4\hsize\epsfbox{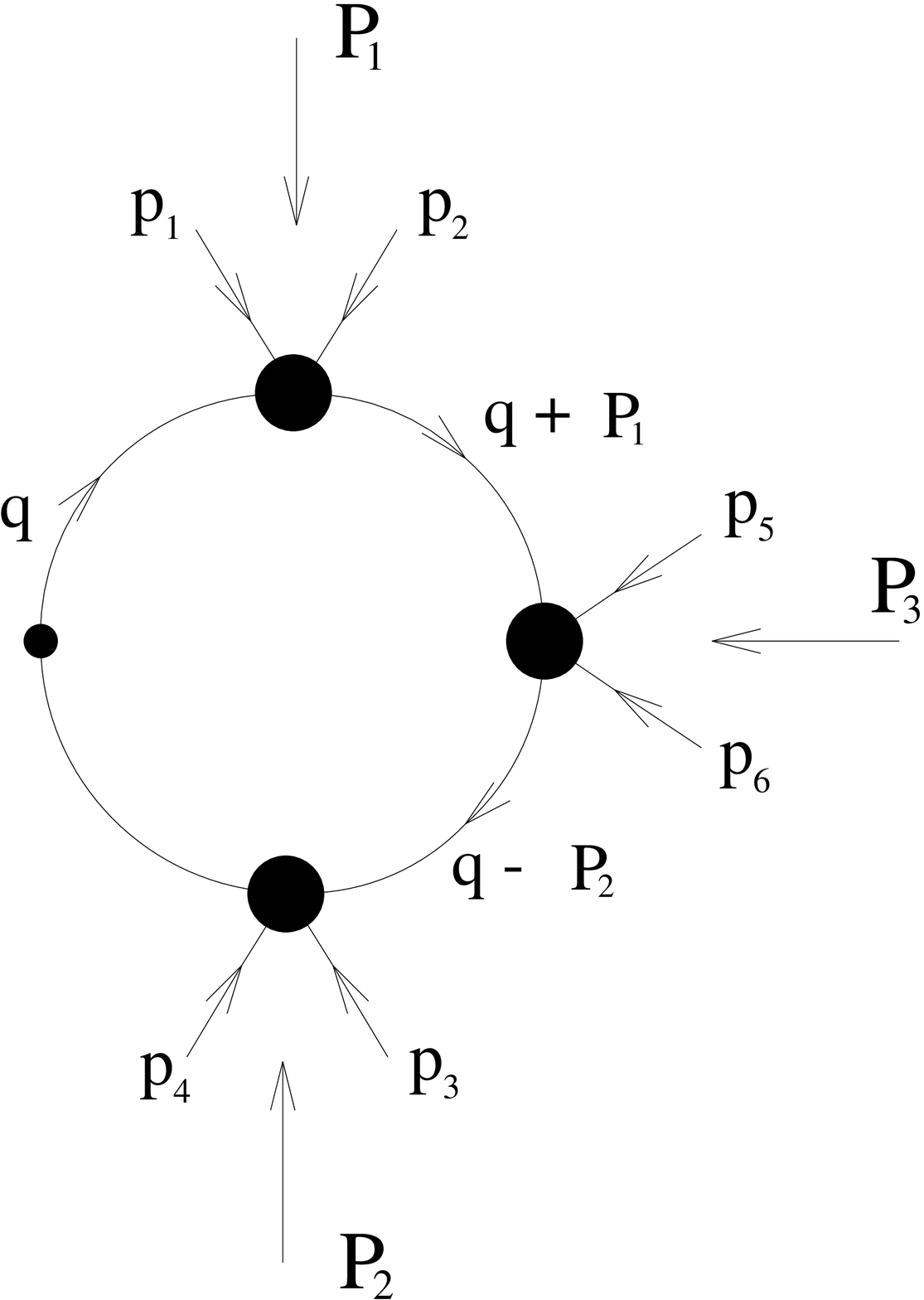}}  
\bigskip
\centerline{\vbox{\noindent {\bf Fig.4.}
One graphical contribution to the flow of the six-point function.
}}
\endinsert
Expanding \yucki\
to $O(p^1)$ we find that, as well as averages that directly follow from \avez\
or sect.5,
 the following averages need to be determined: \eqna\avpi\
$$\eqalignno{& P_2 \big\langle \theta(\q.\P_1)\,  \delta(-\q.{\hat\P_2})
\big\rangle &\avpi a\cr
&  P_1 \big\langle \theta(-\q.\P_2)\, \delta(\q.{\hat\P_1})
\big\rangle &\avpi b\cr
& \big\langle
\theta(\q.\P_1)\,\theta(-\q.\P_2)\,\q.(\P_1-\P_2)\big\rangle\quad,
&\avpi c\cr}$$
where it is always to be understood that these contributions appear summed
over the permutations
${\cal P}(\p_1,\cdots,\p_6)$. First note that \avpi b\ contributes the
same as \avpi a, as can be seen by substituting $\q\mapsto-\q$ in the average
and then permuting $\P_1$ and $\P_2$. (This
permutation simply reorganizes the terms in the sum over permutations.)
Now, substituting $\q\mapsto-\q$ in \avpi a\ gives the equivalent expression
$P_2 \big\langle \theta(-\q.\P_1)\,  \delta(-\q.{\hat\P_2})
\big\rangle$, on using the evenness of the $\delta$-function. Taking the
average of this and \avpi a, and using $\theta(\q.\P_1)+\theta(-\q.\P_1)=1$,
we deduce that \avpi a\ equals $\half P_2\big\langle\delta(\q.{\hat\P_2})
\big\rangle=P_2/4$. This solves for \avpi{a,b}. Using $\theta(-\q.\P_2)=
1-\theta(\q.\P_2)$ in \avpi c\ we obtain the equivalent expression:
$$\big\langle\theta(\q.\P_1)\,\q.\P_1\big\rangle -
\big\langle\theta(\q.\P_1)\,\q.\P_2\big\rangle
-\big\langle\theta(\q.\P_1)\,\theta(\q.\P_2)\,\q.(\P_1-\P_2)\big\rangle\quad.$$
The last term in this expression yields zero by the $\P_1\leftrightarrow\P_2$
symmetry
of the sum over permutations. In the second term we may replace $\P_2$ by
$\P_3$, using permutation symmetry, and hence replace the second term
by $-\half\big\langle\theta(\q.\P_1)\,\q.(\P_2+\P_3)\big\rangle$.
But this is just half
the first term, by momentum conservation. Lastly, the first term is $P_1/4$,
by \avez.

Computing the momentum dependence in this way we find that the
six-point function is, to linearised level:
\eqn\sixx{\Gamma(\p_1,\cdots,\p_6;t)=g_0(t)+g_1(t)\!\!\!
\sum_{{\rm pairs} \{i,j\}} \!\!\! |\p_i+\p_j|\quad,}
where the flow equations for the $g_i$ are to be determined.
Substituting this, \anzf, \anzt, and constants for all $O(p^0)$
higher $n$-point functions, we find the equations again close at the linearised
level. The $O(p^0)$ parts resum into a partial differential equation for
the potential $V_t(\phi)$. The bare locality conditions require now
$\alpha_1(0)
=g_1(0)=0$. Determining the first few couplings at the Wilson fixed
point, we find that $\eta$ can be expressed as the solution of a cubic
with coefficients that depend on $\sigma$, after which non-singular solutions
of the fixed point equation for $V(\phi)$, and perturbations around it,
 can be studied. We
find\foot{To simplify matters, we restricted our search only
to the promising region of $\sigma$ where $\sigma<0$ and $\eta>0$.}
$\eta=.0591$, $\nu=.604$ and $\omega=.540\pm.47i$.
This estimate for $\eta$ has improved slightly compared
with the previous truncation, but the estimates for $\nu$ and $\omega$ are
actually worse.

Finally, let us mention that we have extended the approximation to $O(p^2)$
for the four-point function. The only new average that presents itself is
a term of the form
$\big\langle\theta(\q.\P)\,q^\mu
q^\nu\big\rangle=\delta^{\mu\nu}/6$,
which may be evaluated in a similar way to that of \avpi{b}. We find,
by iteration as before, that the four-point function \anzf\ aquires the new
term
$\alpha_2(t)\left\{p_1^2+p_2^2+p_3^2+p_4^2\right\}$,
otherwise all $n$-point functions have the same form as before.
$\eta$ is now given implicitly
as a function of $\sigma$ through solutions of a complicated sextic.
Unfortunately the sextic seems to have no sensible solutions (i.e. such that
$\sigma<0$ and $\eta>0$)!  If we drop instead to an $O(p^0)$ ansatz for
the six-point function [i.e. set $g_1\equiv0$ in \sixx], we find that the
polynomial bounds the anomalous dimension to be $\eta<0.016$, and therefore
we must obtain worse results than our previous approximations.
We again attribute these negative results to the, in general, poor behaviour
of truncations of the field dependence -- although, as discussed above, more
research is necessary to confirm this conjecture.

\newsec{Conclusions.}
We have shown how the renormalization group for the  effective action
with a sharp momentum cutoff\refs{\wegho}, may be organised by expanding
one-particle irreducible parts in terms of homogeneous
functions of momenta of integer degree i.e. by performing
a momentum scale expansion of the I.R. cutoff Legendre effective action.
A systematic sequence, $M=0,1,2,\cdots$, of $O(p^M)$
approximations then follow. Significantly, the $O(p^0)$ approximation
coincides
with the local potential approximation\refs{\nico} and provides further
justification for the ubiquitous use of the latter\refs{\nico{--}\ui}
by demonstrating how it might be systematically improved.

Note that the momentum scale expansion, although it is unambiguous to
perform, is not an expansion in a small parameter: although the external
momenta are regarded, for the purposes of the expansion,  as small
compared to the cutoff $\Lambda$,
they are in fact set to
values $p\lsim\Lambda$ when the results of the expansion are substituted
into the RHS of the flow equations. It is a trivial but important point
to note that this implies that the expansion is a numerical one which either
converges or 
does not:\foot{Since the equations to be solved are non-linear, it is also
possible that solutions fail to exist beyond a certain $M$.}\
in other words there is no hinterland of
asymptotic convergence, where all of quantum field theory's small parameter
expansions lie.
It is not yet known in general whether the momentum expansion actually
converges
(although low order
results compare favourably with other methods when such
methods are available\refs{\alll}).
We might expect that it generally
does, because it corresponds to an expansion in
an appropriately defined `localness' of the effective Lagrangian, the
expansion in momentum scale $p$  corresponding in position space
(for sharp cutoff) to an
expansion in inverse powers of the average
relative distance $r$ between any two points in a vertex.
Indeed, if
there are large non-local contributions to the effective Lagrangian, then
the description in terms of the chosen fields is probably itself
inappropriate and indicates that other degrees of freedom should be
introduced. This is very different from the cases where truncations of the
field dependence itself could be considered appropriate:
such truncations will only be valid if the field amplitude fluctuations
can be considered small -- but this is only true when
mean field theory is a good approximation, i.e. precisely the regime
where (weak coupling) perturbation theory is valid. In this way,
we can understand why truncations of the field dependence in
truly non-perturbative situations are somewhat
unreliable\refs{\trunc}, while 
 momentum expansions of the effective Lagrangian
(with no other approximation) seem to be so successful\refs{\alll}.

Another trivial but important point is to note that universal quantities,
such as critical exponents, depend, at finite order in the momentum
expansion, on the `shape' of cutoff taken -- whether
it be sharp or smooth of some form $C(p,\Lambda)$. More crucially,
the sharp cutoff
and power-law smooth cutoffs\refs{\deriv} are the only forms of
cutoff known that preserve a field reparametrization invariance of
the flow equations after approximation by momentum expansion.
This is necessary for obtaining, with a given form of cutoff, a unique
value\foot{i.e. universal, in the sense of being independent of the details
of the bare action.}\ for the fields anomalous dimension.

Note that, while Taylor expansions in momenta (a.k.a. derivative expansions)
are possible for any finite width $\epsilon$ of the (smooth) cutoff function,
as $\epsilon\to0$ the dominant terms in
these expansions are positive powers of
 ratios $p/\epsilon$, and diverge. The alternative expansion
in momentum scale must be used in the sharp cutoff limit.
As $\epsilon\to0$, the flow equations become independent of the shape of
cutoff and considerably simplify compared to their
smooth cutoff cousins. As discussed in appendix B, the momentum scale expansion
also appears to be rapidly convergent. There is a price to pay however:
the homogeneous functions of momenta, which appear at each order of the
momentum scale expansion, carry in principle much more information than
Taylor expansions. Therefore it seems that
one `pays' for the better convergence
by absorbing more complication into the expansion at each order. Indeed,
we did not succeed here in computing beyond $O(p^0)$ without truncating the
field dependence.
In the smooth cutoff case, it is clear that the
expansion of the flow equations can be organised
in such a way as to give differential equations only\refs{\deriv} but here
there is not such a simple action on the coefficient functions of the field.
It is quite possible for sharp cutoff, beyond $O(p^0)$, that the resulting
flow equations, if the field dependence is not truncated,
 would only be castable as integro-differential equations.
One also needs (in principle) to check that a given fixed
point behaviour can be reached from a local bare action: otherwise the
behaviour may be the result of non-local physics that does not disappear
as the I.R. cutoff becomes much less than typical low energy scales.

Finally, we noted that the results of the $O(p^0)$ approximation are
well-defined for exceptional momenta,
i.e. the limit $\P_i\to0$ is well-defined for any set of exceptional
momenta --
despite the fact that simply putting $\P_i=0$ results in $\theta(0)$'s
which are ill-defined. In our examples beyond $O(p^0)$, in sect.6, the
limits are again well-defined. Let us mention that in all these cases, the
results obtained are the same as would be obtained by taking the exceptional
case first and then letting $\epsilon\to0$. We do not have a proof that these
properties are enjoyed to all orders in $p^n$. If an ambigous result was
obtained at some $O(p^n)$, in the sense that the limit for exceptional
momenta depended on the direction in momentum space
in which this limit was taken, then
one could presumably proceed by defining these quantities  (if necessary)
by the value obtained when the limit
$\epsilon\to0$ is taken after taking the exceptional case.

\bigbreak\bigskip\bigskip\centerline{{\bf Acknowledgements}}\nobreak
It is a pleasure to thank the SERC/PPARC for financial support through an
Advanced Fellowship.

\vfill\eject

\appendix{A}{Momentum scale expansion to one loop.}
In these two appendices we review some  aspects of model results obtained in
ref.\refs{\erg}. We rearrange them a little, to suit the present purposes.
 Working in four dimensions and with massless\foot{This is the less favourable
but crucial case: the massive case has
better convergence properties.}\ ($\phi\leftrightarrow-\phi$
invariant) $\lambda\phi^4$ theory, one obtains to one-loop that the
one-particle irreducible
four-point function is given by
\eqn\mi{
-\lambda_0^2\sum_{i=2}^4\int_{\Lambda}^{\Lambda_0}\!\!{d^4q\over(2\pi)^4}\
{
\theta(|\q+\P_i|-q)\over
q^2(\q+\P_i)^2}\quad,}
where $\lambda_0$ is the bare coupling, at tree level given by
$\Gamma_{\Lambda_0}[\phi]=\int\!\!d^4\!x\,
\half(\partial\phi)^2+\lambda_0\phi^4/4!$.
The momentum integral is restricted to $\Lambda<q<\Lambda_0$, as indicated.
 $\P_i=\p_1+\p_i$, and $\p_1,\cdots,\p_4$ are the four external momenta.
The above expression
corresponds to the standard $s,t$ and $u$ channel one-loop contributions.
It may be obtained from one iteration of the flow eqn.\alm,  and takes the
form expected -- with a sharp I.R. cutoff
multiplying each propagator -- after a little rearrangement
 (c.f. sect.4 and ref.\refs{\erg}). The momentum expansion of the flow
equations corresponds, at this order in pertubation theory,
to expanding \mi\ directly in small
$\P_i$. We immediately see that a Taylor expansion is not possible, as a
result of the $\theta$-function, whose expansion may be taken from \expIR.
In fact using this, and Taylor expanding the denominator in \mi, one
obtains:
\eqn\mii{
 -{3\lambda_0^2\over(4\pi)^2}
\ln\left({\Lambda_0\over\Lambda}\right) +\lambda_0^2\sum_{i=2}^4\left\{
{P_i\over6}\left({1\over\Lambda}-{1\over\Lambda_0}\right)
+{P_i^3\over720}\left({1\over\Lambda^3}-{1\over\Lambda_0^3}\right)
+O(P_i^5)\right\}\quad.
}
The essential point is that this is an expansion in momentum scale $P_i$,
and not a Taylor expansion in momentum components $P_i^\mu$. In fact, the
even powers of
momentum scale are missing, in this example, to all higher
orders.

\appendix{B}{Convergence in a two loop example.}
In this appendix we go on to investigate the convergence of the momentum
scale expansion in perturbation theory. It is at two loops where one first
has to confront the fact that the momentum scale expansion is not an
expansion in a small parameter. In fact the only contribution to the four-point
function that does not trivially converge, is shown in \fig\fey{ .}.
\midinsert
\centerline{
\epsfxsize=0.3\hsize\epsfbox{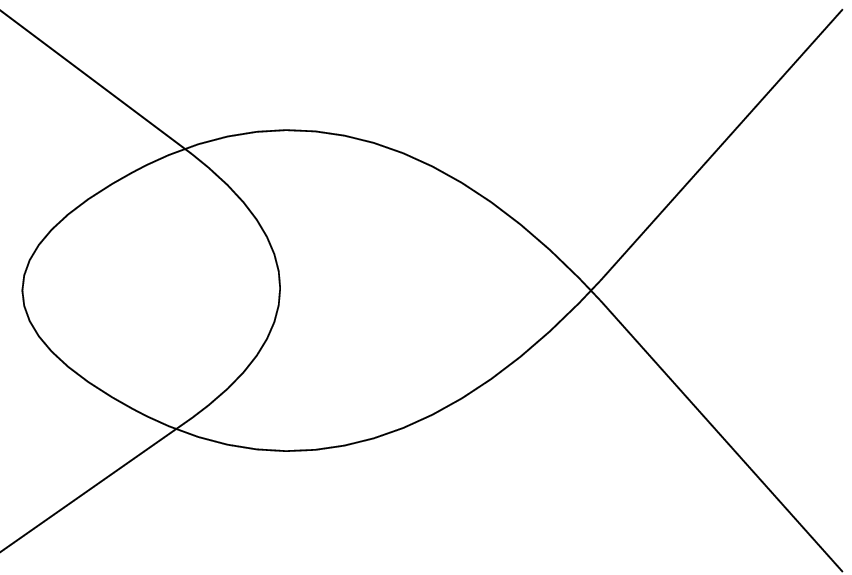}}  
\bigskip
\centerline{\vbox{\noindent {\bf Fig.5.}
The two-loop four-point graph
 whose momentum expansion does not trivially converge.
}}
\endinsert
It arises by iterating \alm\ twice,
and corresponds to the sum of
 two contributions, one where (say) the $t$ channel
contribution from \mi\ has two of its legs tied together by replacing
$P_1$ and $\Lambda$ by $p$,
weighting by the two outer propagators i.e. by $\sim1/p^4$,
 and  then integrating over $\Lambda<p<\Lambda_0$,
and another similar contribution constructed from the one-loop six-point
function. Since the two contributions behave similarly under momentum
expansion, we only investigate the former. Furthermore, we restrict our
attention to the divergent parts, and can therefore set all external
momenta to zero. (The finite parts in fact converge somewhat faster.)
Therefore we are left to investigate a contribution of the form
\eqn\miii{6\lambda_0^3\int^{\Lambda_0}_{\Lambda}\!\!{d^4p\over(2\pi)^4}\
{1\over p^4}\int^{\Lambda_0}_{p}\!\!{d^4q\over(2\pi)^4}\
{\theta(|\q+\p|-q)\over q^2 (\q+\p)^2}\quad.}
In the momentum scale expansion we are to expand the inner one-loop
integral as we did in appendix A. Therefore we can simply substitute the
 result corresponding to \mii. After performing the $p$ integral one obtains
the exact leading divergence as expected\refs{\erg}, while the subleading
divergence is given as a rapidly convergent numerical series:
$$-\beta_2{\lambda_0^3\over (4\pi)^4}\ln{\Lambda_0\over\Lambda},$$
where $\beta_2={1\over\pi}(8+{1\over15}+{9\over2800}+\cdots)$
is a contribution to the two-loop $\beta$-function coefficient.
The ratios of the partial sums to the exact answer $\beta_2=2.568818$, are
$r=.99130,.99956,.99996,\cdots$,
corresponding to the $O(p^M)$
contributions
where $M=1,3,5,\cdots$, and converge to 3sf already at
$O(p^3)$, after which approximately an extra decimal
place in accuracy is added with each new term. We expect that in this way,
the sharp cutoff momentum
scale expansion can be shown to converge at any desired
 order of the loop expansion.

Now we briefly compare with smooth cutoffs. The convergence of the
momentum expansion with power law
smooth cutoffs is rather subtle\refs{\deriv},
in that the polynomial corrections to
the full inverse propagator play a crucial r\^ole in the convergence of the
momentum integrals. Their momentum expansion
 convergence properties will be discussed
elsewhere. Here we consider only a simpler example of an
exponential infrared cutoff $1/q^2\mapsto\theta(q,\Lambda)/q^2$, where
$\theta(q,\Lambda)= 1-\exp(-q^2/\Lambda^2)$, which can directly be
compared with the above example. We might expect at least as good a
convergence of the momentum expansion
 using this cutoff as with power law cutoffs.
The smooth cutoff equivalent of \miii\ is
\eqn\miv{12\lambda_0^3\int^{\Lambda_0}_{\Lambda}\!\!\!\!d\Lambda_1
\int\!\!{d^4p\over(2\pi)^4}\ {\theta(p,\Lambda_1)\over p^4}
{\partial\over\partial\Lambda_1}\theta(p,\Lambda_1)
\int^{\Lambda_0}_{\Lambda_1}\!\!\!\!d\Lambda_2\ I_q\quad,}
where the one-loop integral $I_q$ is given by
$$\eqalign{I_q &=
\int\!\!{d^4q\over(2\pi)^4}\ {\theta(|\q+\p|,\Lambda_2)\over q^2(\q+\p)^2}
{\partial\over\partial\Lambda_2}\theta(q,\Lambda_2)\cr
&= {1\over\Lambda_2}\left({\Lambda_2\over p}\right)^2\left\{1-\exp
\left(-{p^2\over2\Lambda_2^2}\right)\right\}\quad.}$$
The momentum scale expansion of $I_q$ results in convergent momentum integrals
and gives of course the second line expanded as a power series in
$p^2/\Lambda_2^2$. Performing the momentum expansion and the remaining
integrals, one obtains the following closed expression for the
corresponding $\beta_2$:
$$\beta_2=12\sum_{n=1}^\infty {(-1)^{n+1}\over n(n+1)}
\left({1\over2}\right)^n\left[1-\left({1\over2}\right)^{n+1}\right]\quad.$$
The partial sums up to $n=M$ correspond to the $O(p^{2M})$ approximations.
It is clear that this sequence converges.
$\beta_2$ is different from its sharp cutoff value because, as described
above, we have not included
the term that follows from the one-loop six point function.
We can compare ratios (of the partial sums to the full sum) however.
In this case, these are $r=1.18269,.95273,1.01432,\cdots$,
corresponding to the $O(p^M)$ contributions
where $M=2,4,6,\cdots$. We see that convergence is about twice as slow as
for the sharp cutoff momentum scale expansion, in this example.
Slower convergence for smooth cutoffs
is expected, on general grounds\refs{\erg}.

\listrefs

\end